\pdfoutput=1 

\documentclass[twocolumn]{aastex62}

\usepackage{graphicx}
\usepackage[FIGTOPCAP]{subfigure}
\usepackage{amsmath,booktabs,threeparttable,url, bm}
\usepackage{CJKutf8}
\newcommand{\cntext}[1]{\begin{CJK*}{UTF8}{bsmi}#1\end{CJK*}}
\usepackage{float}

\usepackage{natbib}
\received{\today}

\shorttitle{CCSN in binary systems}
\shortauthors{Wang \& Pan 2023}

\graphicspath{{./}{figures/}}

\begin{document}

\title{The Influence of Stellar Rotation in Binary Systems on Core-Collapse Supernova Progenitors and Multi-messenger Signals}

\newcommand*{\NTHUA}{Institute of Astronomy, National Tsing Hua University, Hsinchu 30013, Taiwan}
\newcommand*{\NTHUP}{Department of Physics, National Tsing Hua University, Hsinchu 30013, Taiwan}
\newcommand*{\CICA}{Center for Informatics and Computation in Astronomy, National Tsing Hua University, Hsinchu 30013, Taiwan}
\newcommand*{\CTC}{Center for Theory and Computation, National Tsing Hua University, Hsinchu 30013, Taiwan}
\newcommand*{\NCTS}{Physics Division, National Center for Theoretical Sciences, Taipei 10617, Taiwan}


\author[0009-0003-0630-809X]{Hao-Sheng Wang (\cntext{王皓陞})}
\affiliation{\NTHUA} \affiliation{\NTHUP} \affiliation{\CICA} 

\author[0000-0002-1473-9880]{Kuo-Chuan Pan (\cntext{潘國全})}
\affiliation{\NTHUA} \affiliation{\NTHUP} \affiliation{\CICA}  \affiliation{\CTC} \affiliation{\NCTS}
\email{kuochuan.pan@gapp.nthu.edu.tw}

\begin{abstract}
The detailed structure of core-collapse supernova progenitors is crucial for studying supernova explosion engines and the corresponding multimessenger signals. In this paper, we investigate the influence of stellar rotation on binary systems consisting of a 30 solar mass donor star and a 20 solar mass accretor using the MESA stellar evolution code. 
We find that through mass transfer in binary systems, fast-rotating red- and blue-supergiant progenitors can be formed within a certain range of initial orbital periods, albeit the correlation is not linear.
We also find that even with the same initial mass ratio of the binary system, the resulting final masses of the collapsars, the iron core masses, the compactness parameters, and the final rotational rates can vary widely and are sensitive to the initial orbital periods. For instance, the progenitors with strong convection form a thinner Si-shell and a wider O-shell compared to those in single-star systems. 
In addition, we conduct two-dimensional self-consistent core-collapse supernova simulations with neutrino transport for these rotating progenitors derived from binary stellar evolution. We find that the neutrino and gravitational-wave signatures of these binary progenitors could exhibit significant variations. Progenitors with larger compactness parameters produce more massive proto-neutron stars, have higher mass-accretion rates, and emit brighter neutrino luminosity and louder gravitational emissions. Finally, we observe stellar-mass black hole formation in some of our failed exploding models.  

\end{abstract}


\section{INTRODUCTION}
\label{sec:intro}

For the field of gravitational wave (GW) astronomy and multimessenger astrophysics, the detection of GWs emitted by nearby core-collapse supernovae (CCSN) will be a significant milestone \citep{2021PhRvD.104l2004A, 2021PhRvD.104j2002S}. By gaining a detailed understanding of the gravitational waveform derived from numerical simulations of CCSN, we can enhance GW search pipelines in current observations and contribute to the design of next-generation detectors \citep{2019PhRvD..99f3018R}. Starting from May 24th, 2023, the joined observation run 4 (O4), involving the advanced LIGO, VIRGO, and KAGRA (LVK), has initiated an 18-month period of active observation. With the enhancements of its sensitivities in O4, we are expected to see more frequent and weaker GW events than the previous O3 \citep{2021arXiv211103606T}.

In addition, despite there are still numerical challenges in CCSN theory \citep{2016ARNPS..66..341J, 2020LRCA....6....4M,  2021Natur.589...29B}, it has long been recognized that CCSNe are ideal multimessenger sources that will emit not only multi-wavelength electromagnetism (EM) waves and burst GW signals, but also MeV neutrinos with different flavors \citep{2013ApJ...762..126O, 2020ApJ...901..108V}. Having these multimessenger signals complement each other with their distinct observational characteristics can contribute to the comprehensive understanding of CCSNe. For instance, EM-wave observations can be used to constrain the mass of the CCSN ejecta \citep{1996ApJ...460..408T, 2009ApJ...700..579R}, while gravitational wave and neutrino observations provide insights into the physical properties of the collapsing core \citep{1987PhRvL..58.1490H,2020arXiv201004356A, 2022ApJ...939...13C, 2023arXiv230601919P}.

However, the gravitational waveforms provided from recent CCSN simulations \citep{2013ApJ...768..115O,2017MNRAS.468.2032A,2017ApJ...851...62K,2020ApJ...901..108V,2018ApJ...865...81O,2019ApJ...876L...9R,2021ApJ...923..201A,2022hgwa.bookE..21A,2023PhRvD.107d3008M,2023PhRvD.107j3015V, 2023arXiv230805798B} of typical CCSNe suggest that only galactic events are more likely to be detected by the current LVK detectors. Unfortunately, the galactic CCSN rate is approximately only two events per century \citep{2006Natur.439...45D}.
On the other hand, recent studies have demonstrated that under certain extreme conditions, such as fast-rotating CCSN progenitors, stronger GW signals with specific waveforms can be emitted \citep{2002A&A...393..523D,2017PhRvD..95f3019R,2018MNRAS.475L..91T,2019MNRAS.486.2238A,2019ApJ...878...13P,2020MNRAS.494.4665P,2020MNRAS.493L.138S,2021ApJ...914..140P, 2022MNRAS.510.5535J, 2023arXiv230905161S, 2023arXiv231020411H} and may push the detection limit to extra-galactic.
It is, therefore, natural to investigate the GW emissions from rapid-rotating CCSNe in order to search for GW signals from CCSN during the current O4 or future observation runs. 

Our understanding of rotating CCSN progenitors has significantly improved through a wide range of numerical simulations in the past decade. Several studies have focused on the effects of stellar rotation on the evolution of massive stars \citep{2000ApJ...528..368H, 2000A&A...361..101M, 2012A&A...537A.146E, 2015A&A...581A..15S}. 
However, these studies focused on the rotational effects only during the earlier stages of stellar evolution. It is crucial to examine the influence of stellar rotation on the progenitors of CCSNe in order to gain insights into the later phases of stellar evolution prior to core collapse.

One should note that many of the above-mentioned CCSN simulations are conducted based on the progenitor model introduced by \cite{2007PhR...442..269W} for single-star evolution and have utilized artificial rotational profiles from \cite{1985A&A...146..260E}. However, several literatures suggest that the evolution of rotation during stellar evolution will affect the progenitor's structure. 

For instance, \cite{2015A&A...581A..15S} found that chemically homogeneous evolution (CHE) could greatly enhance mixing in the hydrogen envelope and, therefore, widen the region of hydrogen-burning shell, resulting in a hydrogen-poor envelope in rotating stars. \cite{2020MNRAS.493..518K} explored how internal magnetic braking can deplete the total angular momentum reservoir and lead to a spin-down effect on surface rotation. \cite{2021ApJ...914..105J} found that massive stars within AGN disks could spin up to critical rotation and form compact cores due to mass accretion and tidal forces within AGN disks.
Furthermore, many studies have attempted to incorporate realistic rotation into single-star evolution models. For instance, \cite{2000ApJ...528..368H} and \cite{2000A&A...361..101M}, investigated the effects of different initial masses and rotation rates on massive-star evolution. However, none of the simulation models resulted in the formation of fast-rotating progenitors.

To achieve a more realistic representation of fast-rotating CCSN progenitors, it is essential to consider the fact that the majority of massive stars reside in binary or multi-stellar systems \citep{1976ApJS...30..273A,1978ApJS...36..241A}. For instance, mass accretion from a binary companion commonly leads to the angular momentum transfer from donor to accretor and spin-up of stellar rotation \citep{1981A&A...102...17P}.

In this study, we include binary evolution in the stellar evolution model to consider binary interactions. We then perform multi-dimensional CCSN simulations using these realistic rotating progenitors derived from binary stellar evolution. Furthermore, we present an analysis that combines various aspects of the GW signals produced by our fast-rotating CCSN progenitor models, aiming to assist future efforts in detecting GW signatures emanating from CCSNe.

Moreover, CCSNe are the birthplaces of stellar-mass black holes (BHs). The influence of rotation on the dynamics of BH formation, including the exploration of phenomena such as the Standing Accretion Shock Instability (SASI) and Proto-Neutron Star (PNS) oscillations, has been investigated in \cite{2018MNRAS.477L..80K, 2018ApJ...852...28S, 2021ApJ...914..140P, 2023arXiv230805798B}. In this paper, we present results from some of the progenitors in our simulations, where extreme conditions are encountered during the BH formation process. Our primary objective is to gain insight into how different progenitor structures impact the dynamics of BH formation and the resulting multimessenger signals.

The paper is organized as follows. We describe our simulation codes, numerical methods, and the treatment of physics in Section~\ref{sec:method}. In Section~\ref{sec:binary}, we present the simulation results of our binary stellar evolution of CCSN progenitors. In Section~\ref{sec:ccsn}, we show the results of our CCSN simulations using these binary progenitors. Finally, we summarize our results and conclude in Section~\ref{sec:summary}.

\section{NUMERICAL METHODS} \label{sec:method}

The simulations are divided into two stages: 
First, we use the open-source, one-dimensional stellar evolution code, 
Modules for Experiments in Stellar Astrophysics ({\tt MESA}\footnote{\url{http://mesa.sourceforge.net}} r15140, \citealp{2011ApJS..192....3P, 2013ApJS..208....4P, 2015ApJS..220...15P, 2018ApJS..234...34P, 2019ApJS..243...10P}), 
to conduct binary stellar evolution of massive stars, in which the accretor stars will end up as a CCSN. The corresponding {\tt MESA} inlists and processing scripts are available at {\tt Zenodo}\footnote{\url{https://doi.org/10.5281/zenodo.10429194}}.
Second, we use the final outputs of the {\tt MESA} simulations as the CCSN progenitor models and then perform two-dimensional axisymmetric hydrodynamics simulations with self-consistent neutrino transport, using the {\tt FLASH4}\footnote{\url{https://flash.rochester.edu}} code \citep{2000ApJS..131..273F, 2008PhST..132a4046D} and the Isotropic Diffusion Source Approximation (IDSA, \citealt{2009ApJ...698.1174L}) for the neutrino transport.
Below, we describe each stage of the simulations in detail. 

\begin{deluxetable*}{ccccccccccccc}
\tablewidth{0pt} 
\tablenum{1}
\label{tab:models}
\tablecaption{12 CCSN Progenitor models considered in this study. Different columns represent the following physical properties: model name, the progenitor type when oxygen core forms, sucessful explosion or not, the initial binary orbital period, $P_{\rm int}$, the total mass of the progenitor at core collapse, $M_{\rm SN}$, the progenitor's helium core mass, $M_{\rm He}$, the progenitor's carbon-oxygen core mass, $M_{\rm CO}$, the progenitor's iron core mass, $M_{\rm Fe}$, core-collapse time from $\rho_c = 3\times 10^9$ g cm$^{-3}$ to core bounce, $t_{cc}$, compactness parameter measured at $\rho_c = 10^{11}$ g cm$^{-3}$ with enclosed mass = 1.75 $M_\odot$, $\xi_{1.75}$, central angular velocity when the central density reaches $10^9$~g~cm$^{-3}$, $\omega_{d9}$, the ratio of rotational energy to the potential energy at core bounce, $T/|W|$, the strength of GW bounce signal, $\Delta h_+$, assuming a distance $d=10$~kpc.}

\tablehead{
Model Name & Progenitor Type & Explosion & $P_{\rm int}$ & $M_{\rm SN}$ & $M_{\rm He}$ & $M_{\rm CO}$ &
$M_{\rm Fe}$ & $t_{\rm cc}$ & $\xi_{1.75}$ & $\omega_{\rm d9}$ & $T/|W|$ & $\Delta h_+$ 
\\
 & & & [days] & [$M_\odot$] & [$M_\odot$] & [$M_\odot$] & [$M_\odot$] & [s] & & [rad~s$^{-1}$] & [$10^{-2}$] & [$10^{-21}$]
}
\startdata 
b20\_p05 & RSG & Failed & $5$  & $10.82$ & $10.82$ & $7.39$ & $2.08$ & $0.303$ & $1.77$ & $0.85$ & $2.10$ & $12.43$\\
b20\_p07 & RSG & Failed & $7$  & $15.12$ & $8.67$  & $6.23$ & $1.77$ & $0.364$ & $1.50$ & $0.70$ & $1.68$ & $8.93$\\
b20\_p10 & RSG & Failed & $10$ & $10.92$ & $10.92$ & $6.58$ & $2.13$ & $0.276$ & $2.16$ & $0.74$ & $1.90$ & $15.49$\\
b20\_p15 & BSG & Success & $15$ & $10.60$ & $10.60$ & $7.43$ & $1.54$ & $0.680$ & $0.42$ & $0.78$ & $2.10$ & $9.77$\\
b20\_p19 & BSG & Success & $19$ & $10.77$ & $10.77$ & $7.91$ & $2.29$ & $0.255$ & $2.28$ & $0.76$ & $1.93$ & $16.85$\\
b20\_p20 & RSG & Failed & $20$ & $9.88$  & $9.88$  & $6.81$ & $2.27$ & $0.295$ & $2.07$ & $0.72$ & $1.64$ & $11.55$\\
b20\_p25 & BSG & Success & $25$ & $11.12$ & $11.12$ & $7.85$ & $1.57$ & $0.642$ & $0.59$ & $0.80$ & $2.29$ & $10.93$\\
b20\_p30 & RSG & Failed & $30$ & $11.31$ & $9.26$  & $6.64$ & $1.93$ & $0.332$ & $1.97$ & $0.77$ & $1.95$ & $13.54$\\
b20\_p50 & RSG & Success & $50$ & $11.55$ & $11.55$ & $7.20$ & $2.27$ & $0.320$ & $2.02$ & $0.65$ & $1.66$ & $10.77$\\
b20\_p70 & RSG & Failed & $70$ & $10.62$ & $10.62$ & $7.39$ & $2.21$ & $0.274$ & $2.21$ & $0.75$ & $1.79$ & $15.28$\\
\hline
\hline
s20\_r00 & RSG & Success & -- & $17.59$ & $6.84$  & $4.71$ & $1.66$ & $0.390$ & $1.45$ & $0.00$ & $0.00$ & $0.46$\\
s20\_r05 & RSG & Failed & -- & $10.86$ & $10.86$ & $5.59$ & $1.83$ & $0.340$ & $1.60$ & $0.37$ & $0.59$ & $5.14$\\
\hline
\enddata
\end{deluxetable*}

\subsection{Binary Supernova Progenitors}

We follow the input setup as described in \cite{2021ApJ...923..277R} to conduct coupled massive binary evolution. 
The Ledoux criterion \citep{1947ApJ...105..305L} with a mixing length parameter of 1.5 is used to determine convection. 
We include time-dependent convection as in \cite{2020MNRAS.493.4333R} based on the description in \cite{1969Ap&SS...5..180A}, semiconvection \citep{1983A&A...126..207L}, thermohaline mixing \citep{1980A&A....91..175K}, and the exponential core overshooting \citep{2000A&A...360..952H}. The {\tt approx21\_cr60\_plus\_co56.net} network is used for nuclear reactions. 

The binary system is assumed to be tidally synchronized at the beginning of the simulation, and a rigid rotation at the zero-age main sequence (ZAMS) is used as the initial condition. Rotational controls follow the description in \cite{2000ApJ...528..368H, 2005ApJ...626..350H}, which include dynamical and secular shear instability, Solberg-Hoiland instability, Eddington-Sweet circulation, Goldreich-Schubert-Fricke instability, and the Spruit-Tayler dynamo for the transport of angular momentum.
In addition, when a star approaches its critical rotation, the wind mass-loss rate will be adjusted to ensure subcritical rotation \citep{2021ApJ...923..277R}.
When the effective temperature $T_{\rm eff}>10000$~K, we follow the stellar wind description in \cite{2001A&A...369..574V} for the hydrogen mass fraction $X(H)>0.4$, otherwise, we use the mass-loss rate from \cite{2000A&A...360..227N}. For cold winds with $T_{\rm eff}<10000$~K, we consider the mass-loss rate proposed in \cite{1988A&AS...72..259D}.

We use an implicit scheme in the binary evolution and adopt the mass-transfer rates provided by \cite{1990A&A...236..385K}. 
We consider binary systems that involve a $30 M_\odot$ donor star and a $20 M_\odot$ accretor star with metalicity $Z=0.01$ and with different initial orbital periods (see Table~\ref{tab:models} for considered periods), starting from ZAMS.
Once the mass transfer is finished and the accretor star is detached from the binary system, we continue to evolve the accretor star up to core collapse as a single star for simplicity. 

Table~\ref{tab:models} summarizes the 10 binary simulations included in this work, each with different initial orbital periods. 
In addition, we conduct two single-star evolution simulations, one with a non-rotating model and the other with a rotating model at an initial rotational rate of $50\%$ of the critical rotational rate, for use as reference models. The initial surface velocity of our rotating single-star model is approximately $336.17$ km~s$^{-1}$.  

In this work, the compactness parameter of the stellar core is defined by the formula provided by \cite{2011ApJ...730...70O},
\begin{equation}
\label{eq:com}
\xi_M= \left. \frac{M/M_\odot}{R(M)/1000\ \rm{km}} \right\rvert_{t={\rm bounce}}
\end{equation}
where $M$ is a given enclosed mass ($M=1.75\ M_\odot$ in this work), and $R(M)$ is the radius measured at the corresponding enclosed mass $M$ when central density, $\rho_c$ , reached $10^{11}$ g~cm$^{-3}$. 
\cite{2014ApJ...783...10S} have shown that $\xi_{1.75}$ gives a better discriminate of the explosion characteristics for CCSN progenitors. 
The compactness parameters for our 12 progenitors are provided in Table~\ref{tab:models}.

\subsection{Core-Collapse Supernova Simulations}

We take the final profile of our {\tt MESA} progenitors as the initial condition and then map it into the 2D cylindrical grids in {\tt FLASH} for the CCSN simulations. 
FLASH is a publicly available, multidimensional hydrodynamics code with Adaptive Mesh Refinement (AMR). 
The code setup is similar to what has been reported in \cite{2016ApJ...817...72P, 2018ApJ...857...13P, 2019JPhG...46a4001P, 2021ApJ...914..140P}.
Here, we briefly summarize the setup for the completeness of this paper. 
We use the IDSA scheme for the transport of electron types of neutrinos with 20 energy bins spaced logarithmically from 3 to 300~MeV.
In the IDSA, the distribution of transported neutrinos is decomposed into a free-streaming and a trapped component. The evolution of these two components can be solved separately and linked by a diffusion source term \citep{2009ApJ...698.1174L, 2016ApJ...817...72P}. 
The trapped component is fully solved in multiple dimensions, while the streaming component is assumed to have spherical symmetry \citep{2016ApJ...817...72P, 2018ApJ...857...13P, 2021ApJ...914..140P}. 
$\mu$ and $\tau$ neutrinos are treated by a leakage scheme \citep{2003MNRAS.342..673R}.

We use the new multiple Poisson solver of \cite{2013ApJ...778..181C} for the calculation of self-gravity. We replace the monopole moment of the gravitational potential with an effective General Relativity (GR) potential to approximate GR effects based on the Case A description in \cite{2006A&A...445..273M}.
For the grid setup, we use 2D cylindrical grids with the PARAMESH \citep{2011ascl.soft06009M} library for AMR. The simulation domain includes the inner $10^4$ km of the progenitor. We employ nine levels of refinement. The central $r < 120$ km sphere has the smallest zone width of $0.488$ km, and we reduce the maximum AMR level based on the distance from the PNS center, giving a $0.2^{\circ}-0,4^{\circ}$ effective angular resolution. We use a radial power-law profile for density and velocity in our simulation to approximate the stellar envelope as the outer boundary condition.

We use the nuclear Equation of States (EoS) unit in {\tt FLASH}, which incorporates the finite-temperature EoS routines from \url{stellar.collapse.org} \citep{2010CQGra..27k4103O, 2013ApJ...765...29C}. The Steiner, Fischer, and Hempel (SFHo) EoS is used in this work. The SFHo EoS uses the density-dependent relativistic mean-field interactions of \cite{2010PhRvC..81a5803T}, and the mass-radius is tuned to fit the neutron star radius observation \citep{2013ApJ...765L...5S}.

\section{CCSN progenitors from massive binaries}
\label{sec:binary}

In this section, we present our binary stellar evolution models and compare them to single-star stellar evolution models. Our research primarily focuses on the evolution of the accretor star due to the spin-up mechanism that occurs during the binary mass transfer phase \citep{1981A&A...102...17P}. 
Each model stars from a $20 M_\odot + 30M_\odot$ binary system with a different initial orbital period and assumes that they are synchronized by the tidal locking effect. In addition, two single-star models with and without rotation are used as reference models.
All models reach the core collapse stage, which is characterized by a central density exceeding $10^{9.5}$~g~cm$^{-3}$. After reaching this stage, the FLASH code is used to handle the dynamical evolution of a CCSN.
Note that we terminate some of these progenitor simulations in MESA slightly earlier than the stopping criteria mentioned above if the time step reaches the minimum time step due to vigorous nuclear reaction around the iron core. 
Some physical properties of the CCSN progenitor systems, such as the total mass of the progenitor at the core collapse, the iron core mass of the progenitor, the compactness parameter, rotational speed, and the strength of the bounce signal, are summarised in Table \ref{tab:models}.

\begin{figure}
\begin{center}
	\plotone{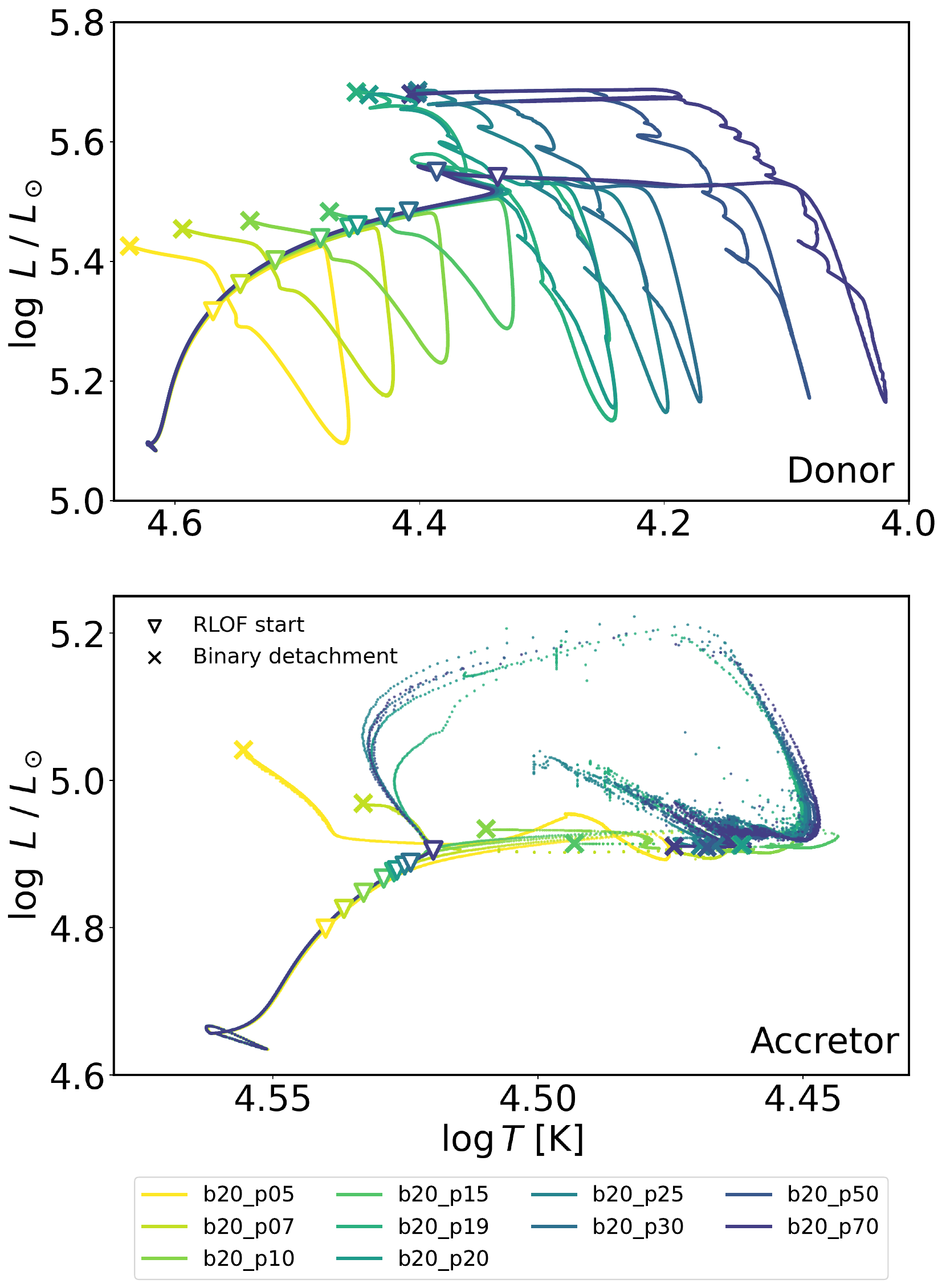}
	\caption{\label{fig:hr_d}
	The HR diagrams illustrate the binary evolution of the donor stars (upper) and accretor stars (lower) from ZAMS to the detachment of the binary systems. Different colors represent evolutionary tracks from different models in Table~\ref{tab:models}. The triangle and star symbols represent the starting point of RLOF and binary detachment, respectively. }
\end{center}
\end{figure}

Figure~\ref{fig:hr_d} illustrates the binary evolution of the donor stars (upper) and accretor stars (lower) from ZAMS (from the lower-left corner) to the point of binary detachment (star symbols). The triangle symbols for each line represent the starting point of Roche Lobe Overflow (RLOF).

For models with initial orbital periods shorter than 15 days, mass transfer begins (the triangles in Figure~\ref{fig:hr_d}) during the early main sequence phase. These donor stars have experienced a significant luminosity drop before they turn into supergiants (see the upper panel in Figure~\ref{fig:hr_d}).
Models with longer initial periods initiate mass transfer during the later phase of the main sequence, resulting in the luminosity drop occurring after the donor stars have evolved into supergiants. In both cases, the steep fall in the donor's luminosity is attributed to the rapid mass transfer, preventing the core from generating enough energy to maintain expansion, as described in \cite{2022A&A...657A.116P}.

In the lower panel of Figure~\ref{fig:hr_d}, we show the evolutionary tracks of the accretor stars on the HR diagram, which vary depending on the binary period. Accretors that undergo binary interaction while their donors are in the early phase of the main sequence exhibit minimal luminosity changes during mass transfer. In contrast, accretors in systems where mass transfer begins while their donor stars are in the later phase of the main sequence experience a significant increase in luminosity ($\log$ L/L$_\odot \sim 4.9$, $\log$ T $\sim 4.52$ K) due to their radius expansion, and their accretion evolution track aligns with that described in \cite{2021ApJ...923..277R}.

\begin{figure}
\begin{center}
	\plotone{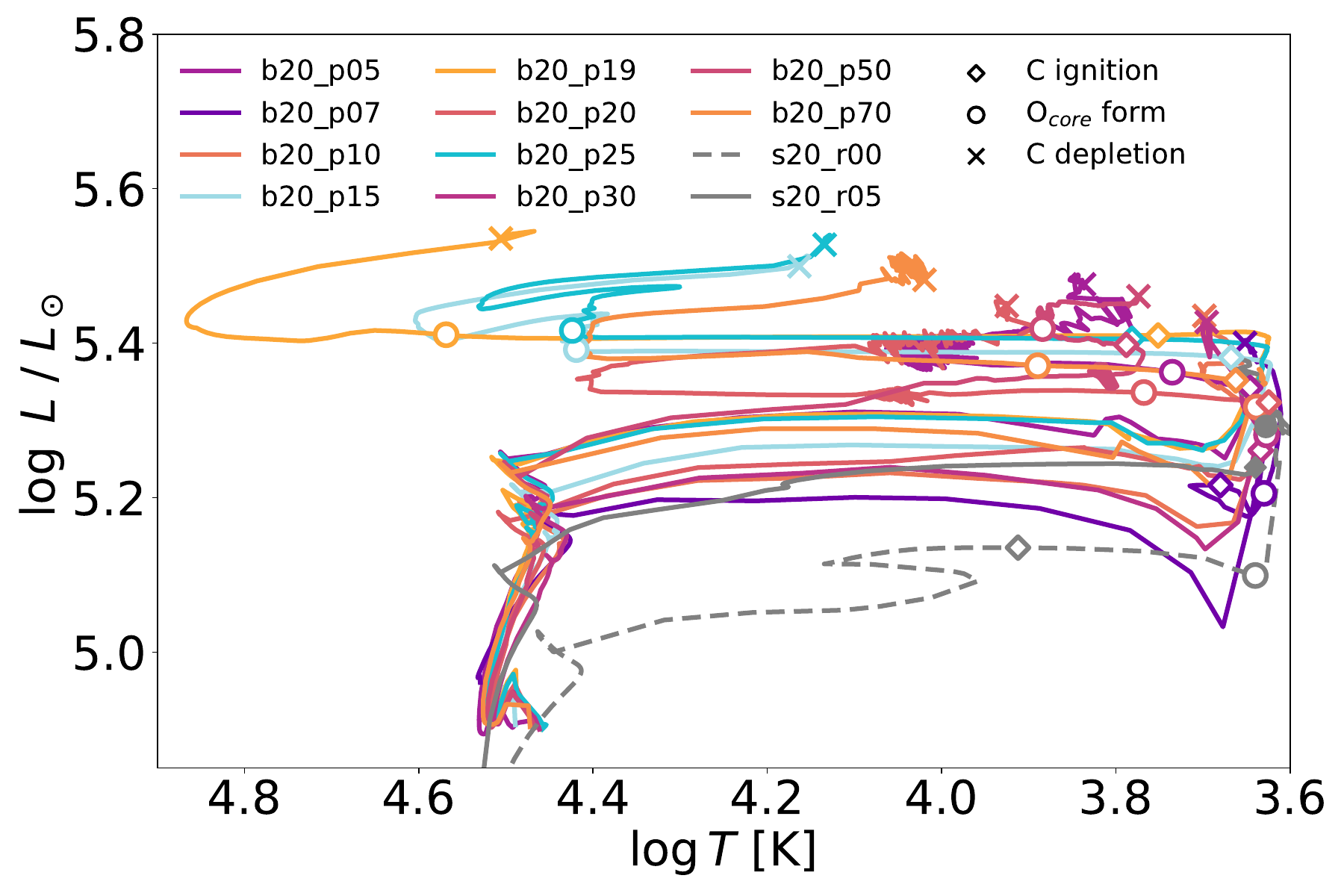}
	\caption{\label{fig:hr_a}
    Evolutionary tracks of accretor stars from binary detachment to carbon depletion stage in the HR diagram. Different colors represent evolutionary tracks from different models in Table~\ref{tab:models}. Markers are used to indicate conditions at carbon ignition (diamonds), oxygen-core formation (circles), and carbon depletion (crosses).}
\end{center}
\end{figure}

After binary detachment, the evolution tracks indicate an absence of a correlation between the initial binary period and the resultant progenitor structure. Figure~\ref{fig:hr_a} describes the following evolutionary tracks of the accretor stars from binary detachment to the carbon depletion stage. To improve the clarity of our presentation, we chose to utilize an alternative physical property, the compactness parameter $\xi_{1.75}$ (see Equation~\ref{eq:com} and Table~\ref{tab:models}), for organizing the line colors. 
The compactness parameter is commonly used to represent the core structure of a CCSN progenitor. A more detailed discussion on why the compactness parameter is better to present our binary progenitor models will be shown in Section~\ref{sec:ccsn}.

Note that all accretors remain in the main sequence after binary detachment. Furthermore, we have omitted the evolutionary tracks beyond carbon depletion in the HR diagram because the luminosity and effective temperature of these accretor stars undergo significant variations during the final stages of their evolution. To maintain clarity, we have chosen not to include them in the HR diagram.

Regarding stellar evolution in the post-detachment (Figure.~\ref{fig:hr_a}),  we observe that the evolutionary tracks exhibit chaotic behavior with no discernible correlation with the initial period. 
In addition, we note that some accretors undergo dramatic contractions and turn bluer after carbon ignition. Other accretors without strong contractions remain red supergiants. 
This variation has also been discussed in previous research \citep{1969ApJ...155..935S,2000ARA&A..38..143M,2021A&A...652A.137E} that HR diagram tracks are sensitive to factors such as nuclear reaction rates, internal mixing, and mass loss. 
In Figure~\ref{fig:hr_a}, the accretors of models {\tt b20\_p15}, {\tt b20\_p19} and {\tt b20\_p25} reach high effective temperatures exceeding  $10^{4.4}$K and have briefly transitioned into blue supergiants (BSG) when the oxygen core formed, while seven accretors do not undergo significant contraction and maintain lower effective temperatures.

It is well-known that the progenitor of SN 1987A was a BSG \citep{1987A&A...177L...1W, 1987ApJ...320..602K}. Two mainstream scenarios to explain the BSG progenitor of SN~1987A are the merger origin \citep{1989ARA&A..27..629A} and the binary origin \citep{1992PASP..104..717P}. We find that the evolution tracks of our BSG models are similar to the potential binary progenitor track reported from the previous study (see Figure~4 in \citealp{1992PASP..104..717P}). Our binary models provide support for the potential binary nature of the SN~1987A progenitor, though our models did not aim to reproduce the SN~1987A progenitor. 

\begin{figure}
\begin{center}
	\plotone{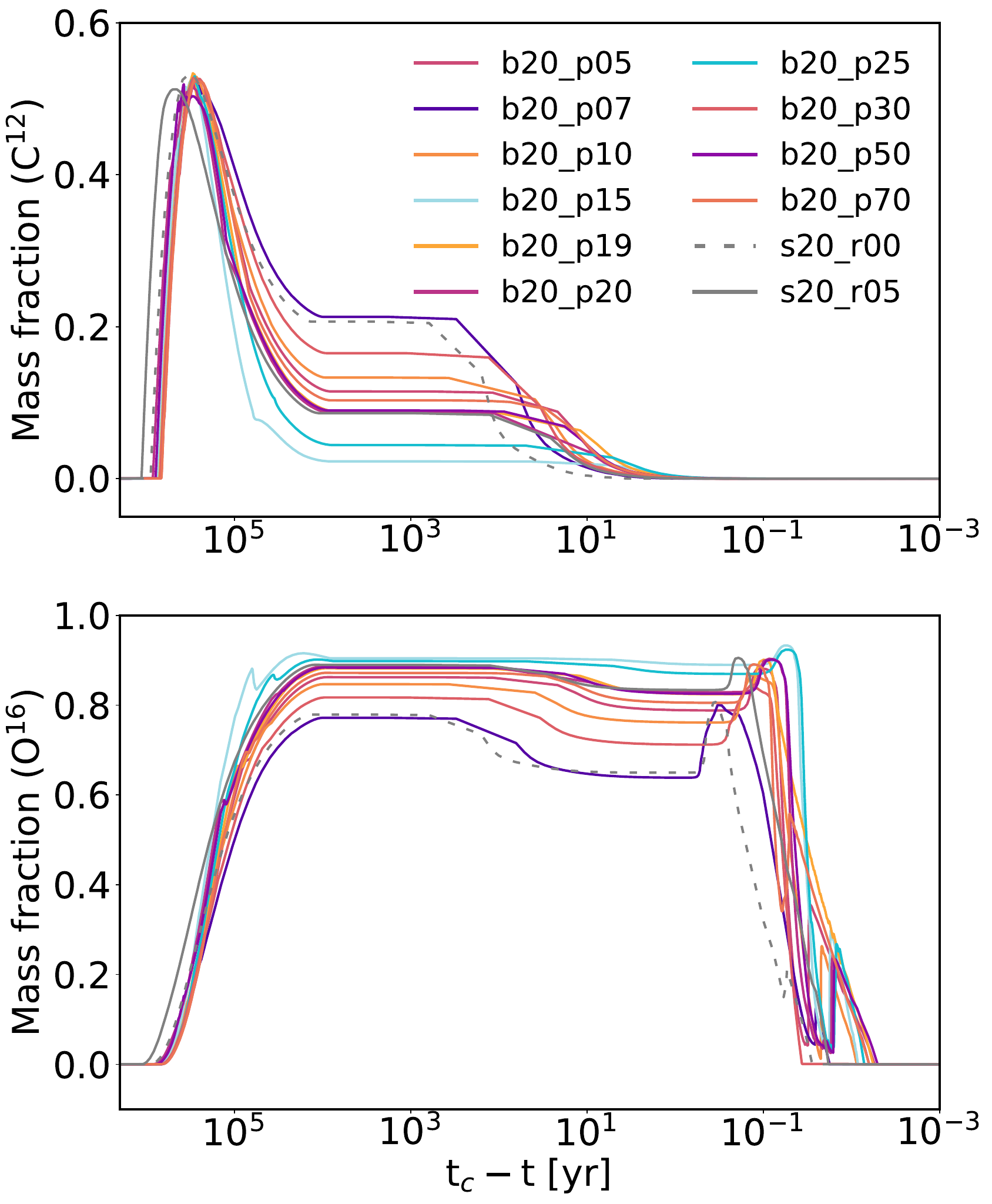}
	\caption{\label{fig:co}
	The mass fraction evolution of C$^{12}$ and O$^{16}$. We line up different models by setting the x-axis as the time before the progenitor collapses.} 
\end{center}
\end{figure}

In addition, numerous studies have emphasized the critical role of carbon-oxygen cores in shaping CCSN progenitor core structures \citep{2018ApJ...860...93S,2021A&A...645A...5S,2022ApJS..262...26L}. 
We investigate the evolution of the central mass fraction to explain the causes of strong contractions during carbon ignition, specifically examining the changes in central carbon (C$^{12}$) and oxygen (O$^{16}$) fractions at various stages. Figure~\ref{fig:co} reveals that models {\tt b20\_p15} and {\tt b20\_p25} (blue lines) contain significantly lower carbon fractions but higher oxygen fractions compared to other models.
This can be understood by the fact that strong convection during the helium-burning stage could produce more $\alpha$-particle for converting carbon into oxygen \citep{2021A&A...645A...5S}. 
On the other hand, the non-rotating single-star model, {\tt s20\_r00}, and the binary progenitor, {\tt b20\_p07}, have the weakest convection among these models, resulting in larger carbon core abundance and less oxygen core abundance. 
Other binary progenitor models fall within the spectrum of these extreme progenitor models, indicating that the structure of CCSN progenitors is significantly influenced by the strength of core convection after leaving the main-sequence stage. 

We note that, while the central density and central temperature for all models at the same stage are similar, the differences in carbon and oxygen fractions can be attributed to variations in convection efficiency, which in turn, affects the nuclear burning rate after the main-sequence stage.

\begin{figure*}
	\plotone{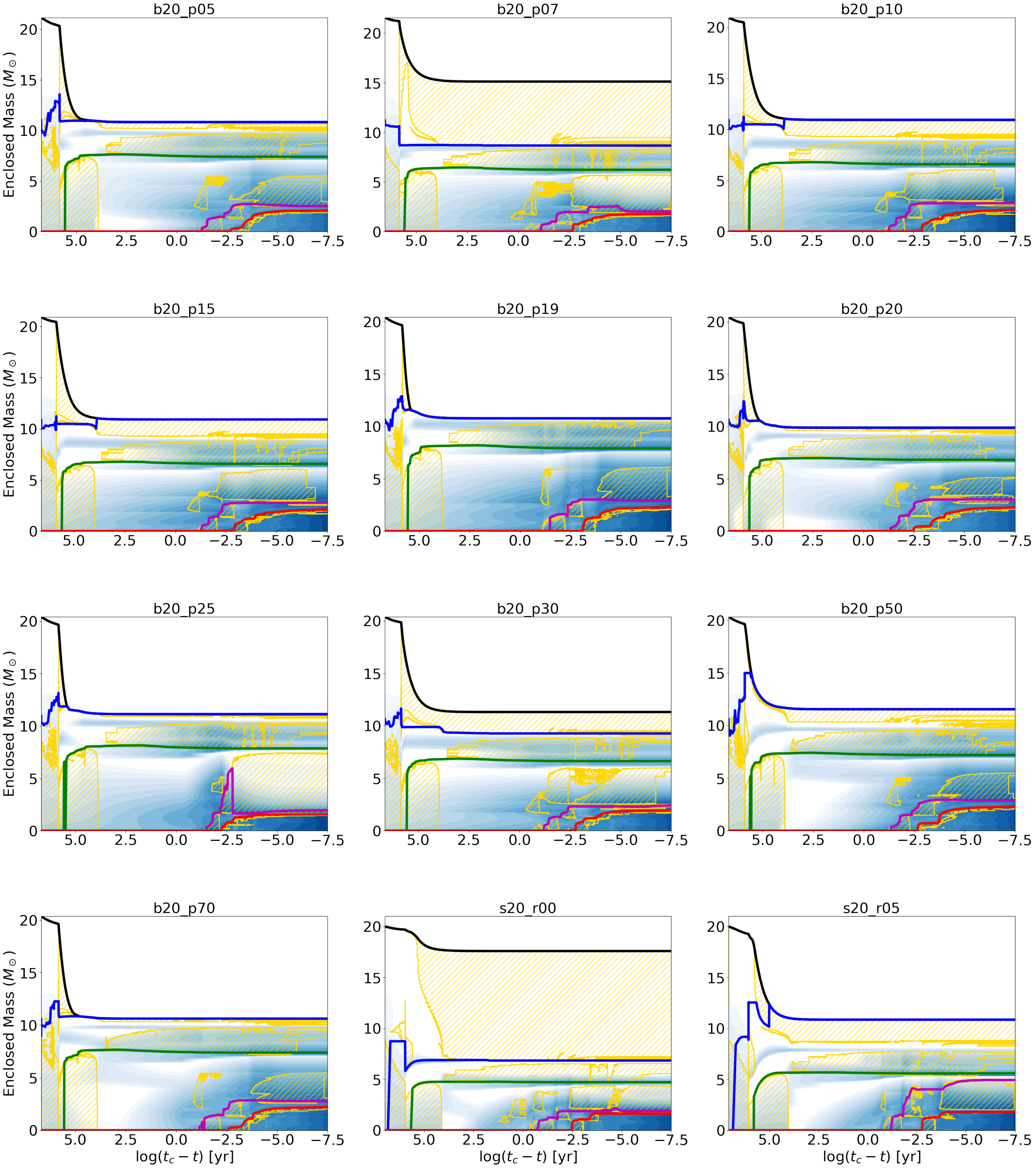}
	\caption{\label{fig:kipp}
	The Kippenhahn diagrams for different models (see Table~\ref{tab:models}) from binary detachment (or at ZAMS for single star models) to core collapse. In each panel, different solid lines represent the boundaries of various shells, such as hydrogen (black), helium (blue), oxygen (green), silicon (purple), and iron (red). The yellow-shaded regions represent the convective region. The blue contour map indicates the nuclear energy of burning.}
\end{figure*}
\begin{figure*}
	\plotone{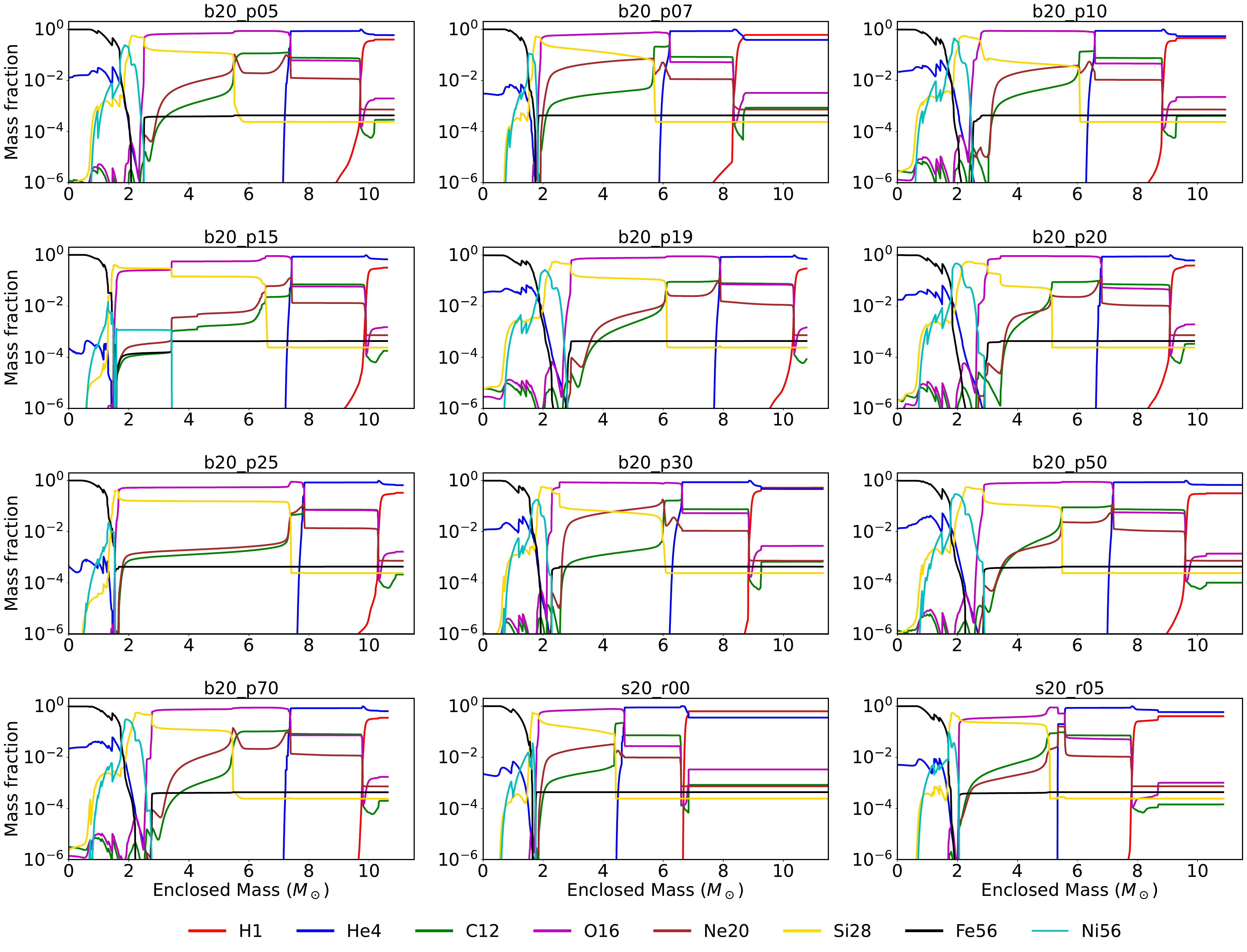}
	\caption{\label{fig:abun}
	The abundance distribution for different CCSN progenitors during core collapse when the central density has reached $\rho_{\rm{c}}=10^9$~g~cm$^{-3}$. Different colors represent the mass fraction of different elements based on the mass coordinate.}
\end{figure*}

Figures \ref{fig:kipp} and \ref{fig:abun} demonstrate the structure evolution after binary detachment and the progenitor abundance at the collapse for different models in Table~\ref{tab:models}. 
Figure~\ref{fig:kipp} shows the Kippenhahn diagrams of our considered models from the point of binary detachment (or at ZAMS for single stars) up to core collapse.
We note that the binary progenitors gain extra mass and angular momentum during the mass-transfer stage, leading to a total mass slightly exceeding $20 M_\odot$ at binary detachment. 
After detachment, both the single-star and binary rotating progenitors experience significant mass loss and lose their hydrogen envelope due to stellar winds during helium burning (see Figure~\ref{fig:kipp}). Binary rotating progenitors, in particular, lose approximately half of their total mass during the helium-burning stage, resulting in them becoming CCSN progenitors with an approximate mass of 10 M$_\odot$ (see Table~\ref{tab:models}).

Meanwhile, rotational effects could enhance the convection and, therefore, enlarge the chemical mixing in the hydrogen-burning shell, resulting in a larger helium shell \citep{2015A&A...581A..15S}. Despite the absence of a hydrogen shell for rotating progenitors in Figure~\ref{fig:kipp}, we do not expect that all progenitors would result in Type I supernovae. In Figure~\ref{fig:abun}, we observe that the hydrogen surface fraction remains similar to that of helium. Such compositions are conducive to the formation of Type IIb supernovae \citep{1994ApJ...429..300W}. 
Furthermore, the carbon-oxygen core masses (see Table~\ref{tab:models}) are enlarged due to strong mixing facilitated by convection, meridional circulation, and turbulent shear in rotating CCSN progenitors \cite{2017hsn..book..513L}.

As we mentioned above, models {\tt b20\_p15} and {\tt b20\_p25} exhibit strong convection during the stellar evolution stage, which will affect the subsequent evolution and make significant differences in the structure of CCSN progenitors. Notably, model {\tt b20\_p25} is the model with the strongest convection during helium burning. In Figure \ref{fig:kipp}, we can discern a reduction in silicon core mass in the model {\tt b20\_p25}. The strong convection around the oxygen and silicon shell interface leads to its mixing with oxygen and the formation of a mixed oxygen-silicon shell outside the iron core. 
Similar effects have been observed in other models of massive stellar evolution \citep{2019ApJ...881...16Y,2020ApJ...890...94Y,2021A&A...656A..58L,2021ApJ...908...44Y}, and these are evident in our analysis of progenitor abundances (see Figure~\ref{fig:abun}). 

For models containing weaker convection, such as {\tt b20\_p10}, {\tt b20\_p20} and {\tt b20\_p50}, the inner convective regions decrease as the progenitors approach collapse. These models form distinct boundaries between the oxygen and silicon shells in Figure~\ref{fig:abun}. 

Among our models, the final rotational energy and speed of our CCSN progenitors depend on several factors, such as the angular momentum transfer from its binary companion, the core contractions, and convection efficiency.
We have observed that binary progenitors exhibit significantly higher $T/|W|$ values than the progenitor from the rotating single-star model (model {\tt s20\_r05}), suggesting that these binary progenitors have a high chance to emit strong GW signals \citep{2017PhRvD..95f3019R}.

\begin{figure}
	\plotone{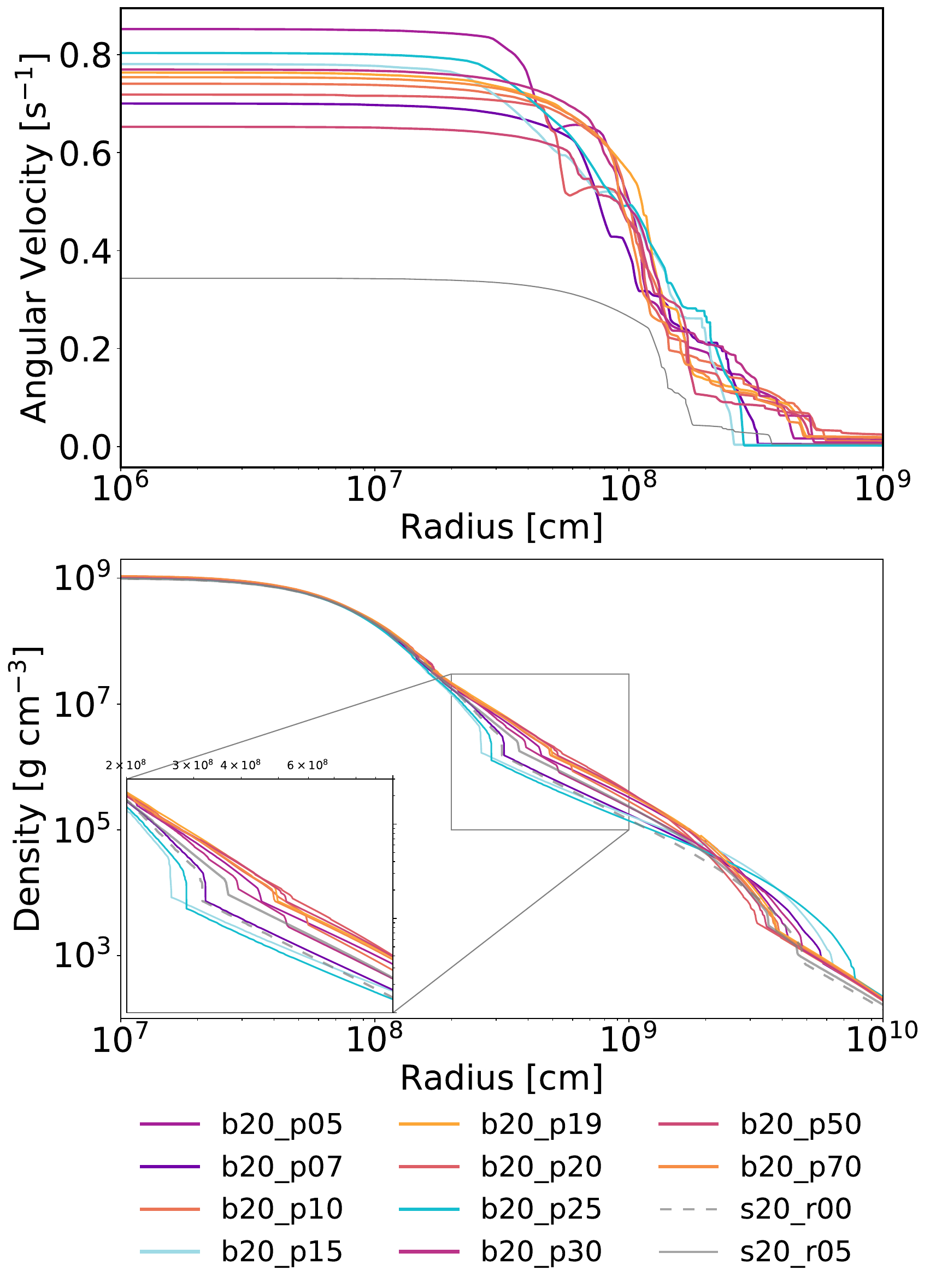}
	\caption{\label{fig:prog}
	Radial distributions of angular velocities and densities. Different lines represent the distribution of different CCSN progenitors when the central density $\rho_{\rm{c}}=10^9$ g cm$^{-3}$.}
\end{figure}

It is also well-known that the structure of CCSN progenitors, including factors such as the compactness parameters, the density gradients, and the iron core sizes, deeply influences the explodability of a supernova model \citep{2014ApJ...783...10S,2016ApJ...821...38S,2018ApJ...860...93S,2021Natur.589...29B}. 
To investigate the inner structure of CCSN progenitors, Figure~\ref{fig:prog} presents the profiles of the angular velocity and the density distribution of the cores of the CCSN progenitors, recorded at the onset of the central density $\rho_{\rm{c}}=10^9\ g\ cm^{-3}$. 
First, it is important to note that there is no linear relationship between the initial orbital period and the final angular velocity, as binary systems tend to form fast-rotating progenitors when compared to single-star progenitors. 
Secondly, within the radii of $10^8$ and $10^9$ cm, strong convection models during the helium-burning stage and the oxygen-silicon shell mergers result in lower compactness parameters and smaller iron cores due to a lower abundance of silicon in their core composition, this behavior supports the idea in \cite{2014ApJ...783...10S} that the large density gradient at iron core surface lead to smaller compactness parameter. 

\begin{figure}
	\plotone{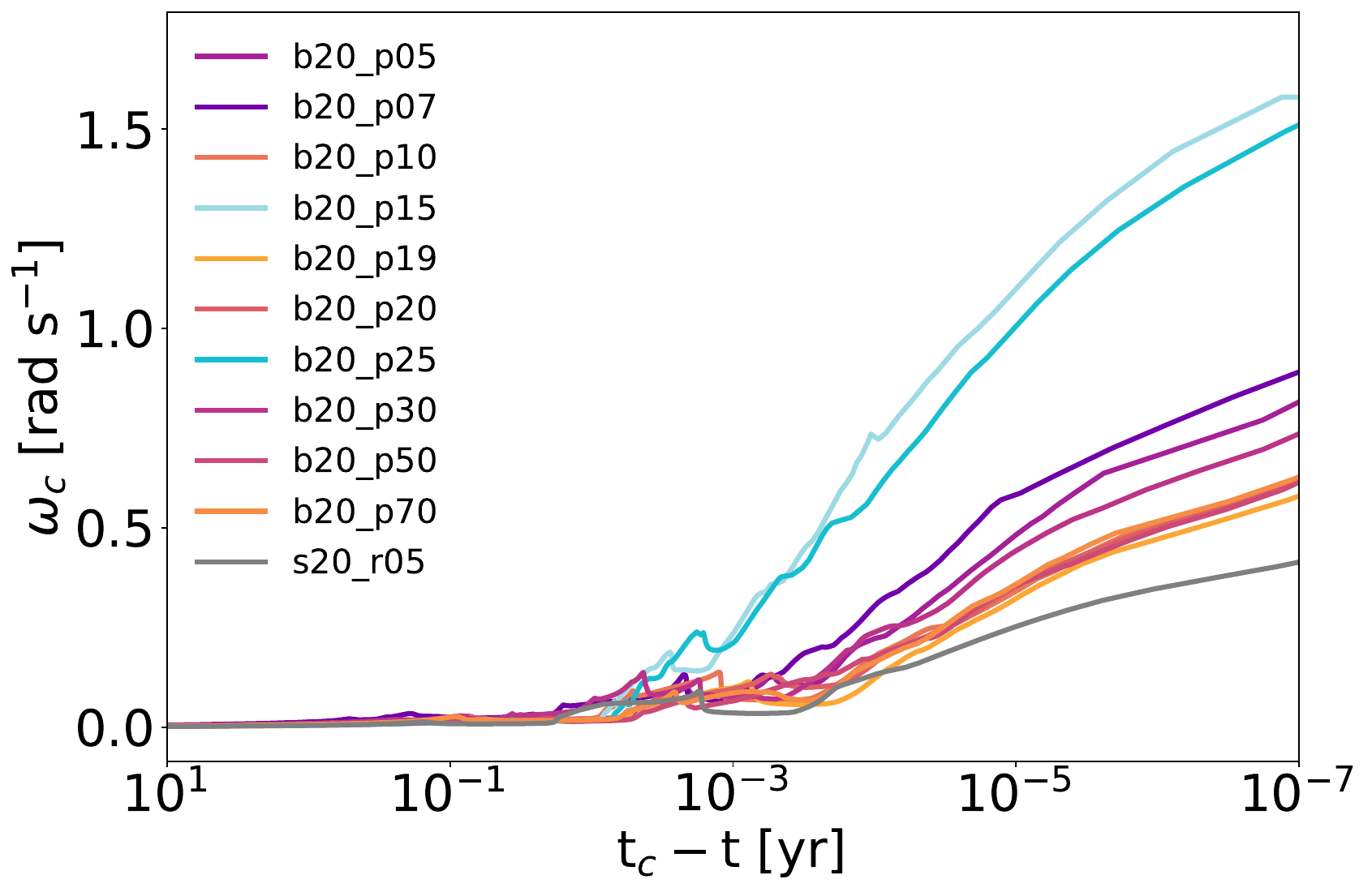}
	\caption{\label{fig:wevo}
	The time evolution of central angular velocity of our considered binary and single-star rotating models. We line up different models by setting the x-axis as the time before the progenitor collapses.}
\end{figure}

Figure~\ref{fig:wevo} illustrates the evolution of central angular velocity ($\omega_c$) in rotating progenitors until the core collapses. During the earlier evolution stage, we do not find significant differences in angular velocity among these models. However, notable distinctions become apparent after silicon ignition. In comparison to the models without oxygen-silicon shell merger, both models {\tt b20\_p15} and {\tt b20\_p25} exhibit significantly faster rotation in their progenitors. 
In addition, \cite{2023ApJS..264...45F} conducted a wide parameter space of population synthesis. They concluded that the final rotational rates of the secondary star are more sensitive to the mass of the donor star than the initial binary period, which is consistent with the non-monotonic relationship of the final central angular velocity in Figure~\ref{fig:wevo}.

From the lower panel of Figure~\ref{fig:prog}, both models {\tt b20\_p15} and {\tt b20\_p25} contain steep composition gradient, which may lower the angular momentum transport efficiency and then end up with a faster-rotating core. The result is consistent with previous research suggested by \cite{2019MNRAS.488.4338M}. Core angular momentum transport efficiency deeply affects the core rotation, and efficient angular momentum transport results in a slower rotating core, and vice versa. 

\section{CCSN in massive binaries and their multimessenger signals}
\label{sec:ccsn}

We conduct 12 CCSN simulations using these 10 binary progenitors and 2 single-star progenitors to investigate the impact of rotation from binary evolution on core-collapse supernovae (see Table~\ref{tab:models}). 
Among these simulations, five cases resulted in successful explosions, 
including four binary progenitors (models {\tt b20\_p15}, {\tt b20\_p19}, 
{\tt b20\_p25}, and {\tt b20\_p50}) and one single-star progenitor (model {\tt s20\_r00}). In the below subsections, We describe the detailed dynamical evolution of our CCSN simulations and their multi-messenger signatures.

\subsection{Dynamical Evolution}
\label{sec:dyna}

\begin{figure}
	\plotone{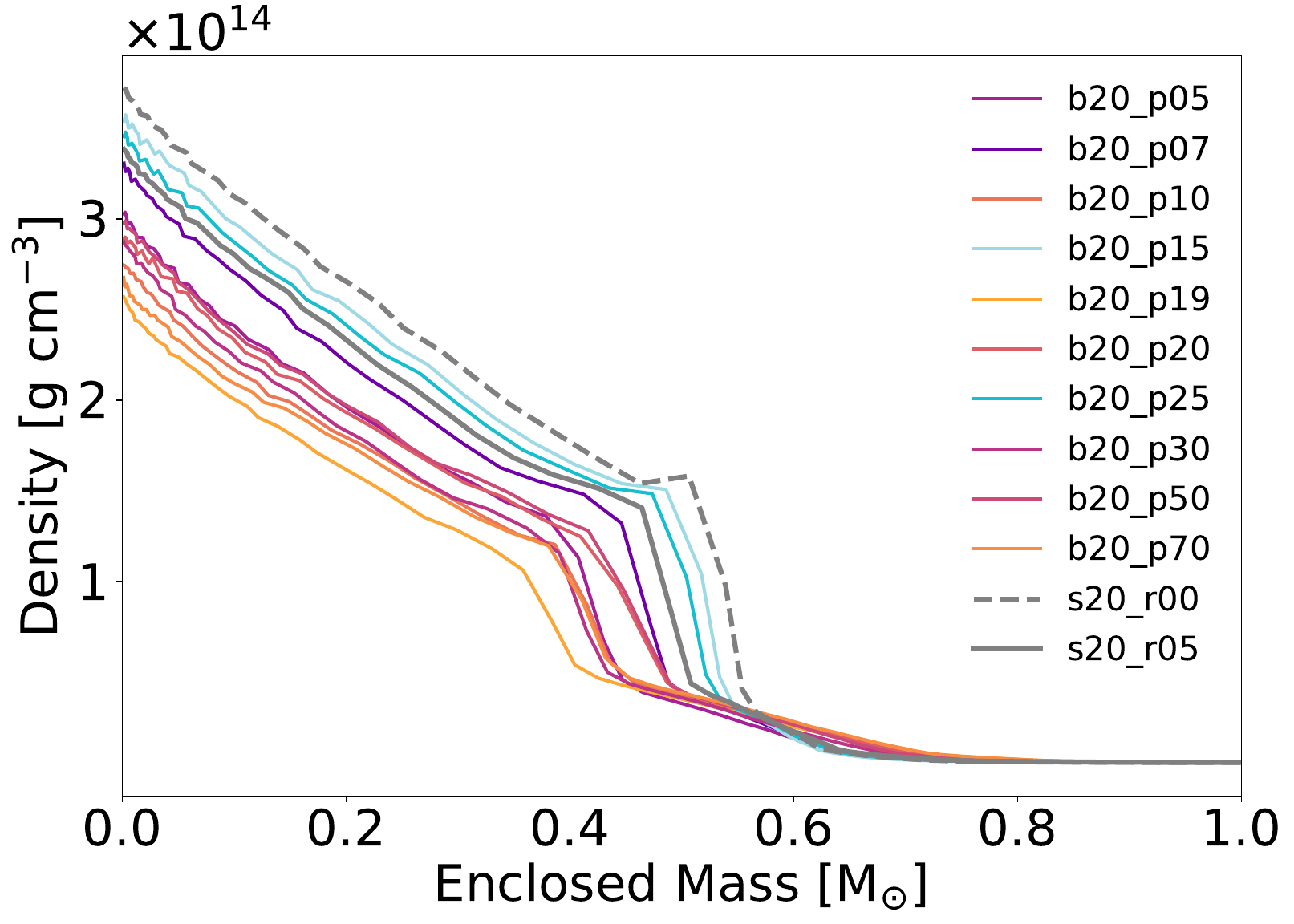}
	\caption{\label{fig:bc_dens}
        The density distribution in the mass coordinates at core bounce. Different colors represent different progenitor models in Table~\ref{tab:models}.} 
\end{figure}

Figure \ref{fig:bc_dens} illustrates the density profiles of our models at core bounce. 
Notably, the progenitors with lighter oxygen and carbon shells due to stronger shell convection ({\tt b20\_p15} and {\tt b20\_p25}), as well as the single-star models, exhibit higher bounce densities compared to the other models. 
This difference in bounce density can be attributed to the compactness parameter (see Equation~\ref{eq:com}), 
with smaller values resulting in longer collapsing times (from $\rho_c = 3 \times 10^9$~g~cm$^{-3}$ to bounce; see Table~\ref{tab:models}) and subsequently higher bounce densities. 
Conversely, models with larger compactness parameters display lower bounce densities. 
This behavior is consistent with the correlation found in \cite{2002RvMP...74.1015W}, where larger core masses correspond to lower central densities at a given central temperature. 
In addition, it's also noteworthy that fast-rotating progenitors yield lower bounce densities due to the influence of centrifugal forces \citep{1994ApJ...434..268Y}.

\begin{figure}
	\plotone{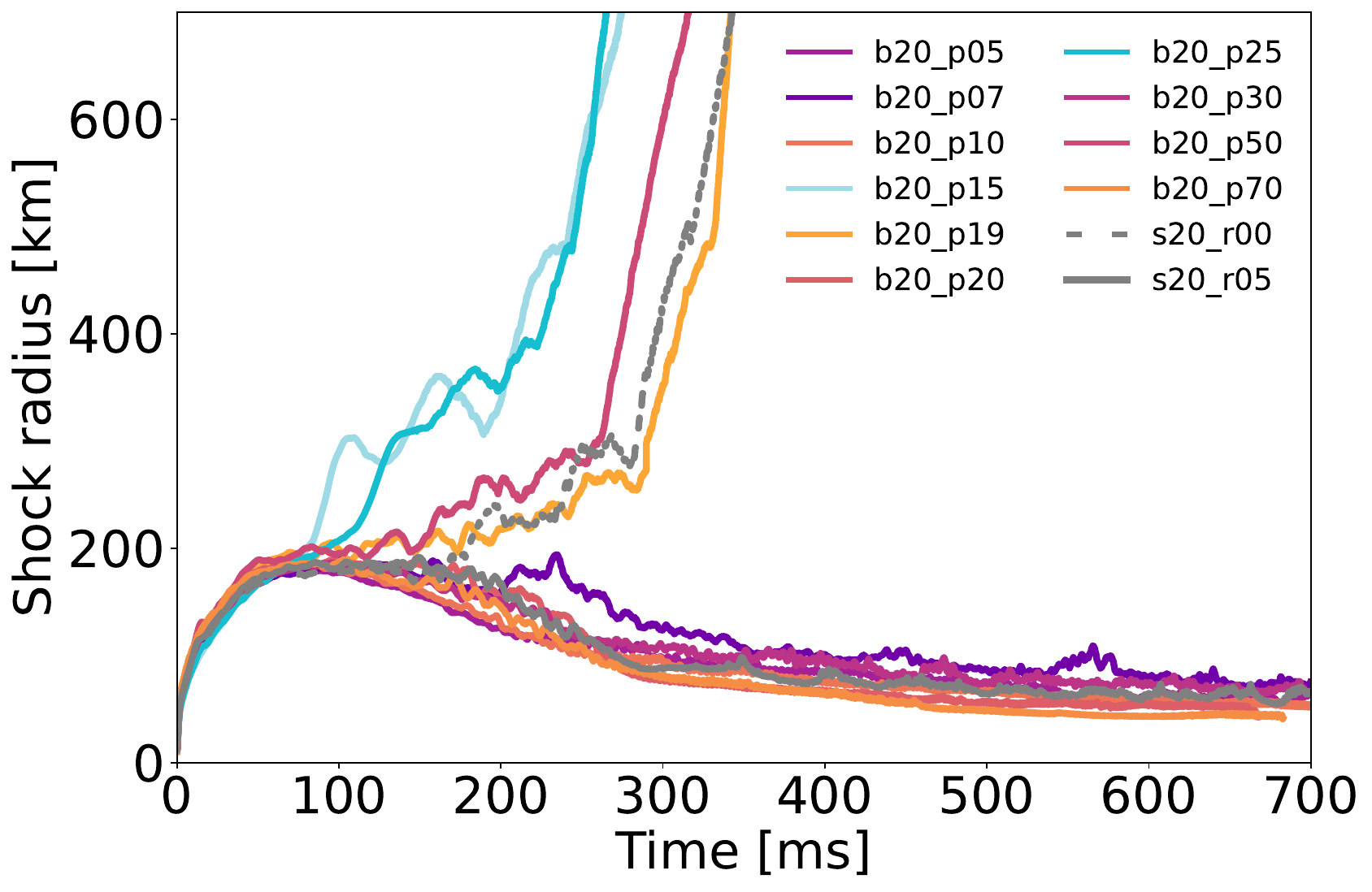}
	\caption{\label{fig:rshock}
	Time evolution of the averaged shock radii for our considered models (see Table~\ref{tab:models}). The colored lines represent the evolutions for different binary progenitor models. while the grey lines represent the shock radius evolution for the single-star models. Specifically, the dashed gray line represents the non-rotating single-star model, and the solid grey line shows the rotating single-star model.}
\end{figure}

Figure~\ref{fig:rshock} shows the time evolution of the averaged shock radius. 
Among these 12 models, five of them experience successful explosions within 300~ms postbounce.
The least compact models ({\tt b20\_p15} and {\tt b20\_p25}) have an early explosion time ($\sim 250$~ms) compared to other exploding models due to their low compactness parameters \citep{2014ApJ...783...10S}. 
Following that, the higher compactness parameter models {\tt b20\_p19}, {\tt b20\_p50}, and {\tt s20\_r00} explode at around $\sim 300$~ms postbounce due to strong neutrino accretion luminosity.
Overall, CCSN simulations with rotations are harder to explode in 2D due to the suppression of convection in 2D \citep{2019ApJ...878...13P}, explaining the reason that only the non-rotating progenitor explodes successfully in our single-star progenitor models.

\begin{figure}
	\plotone{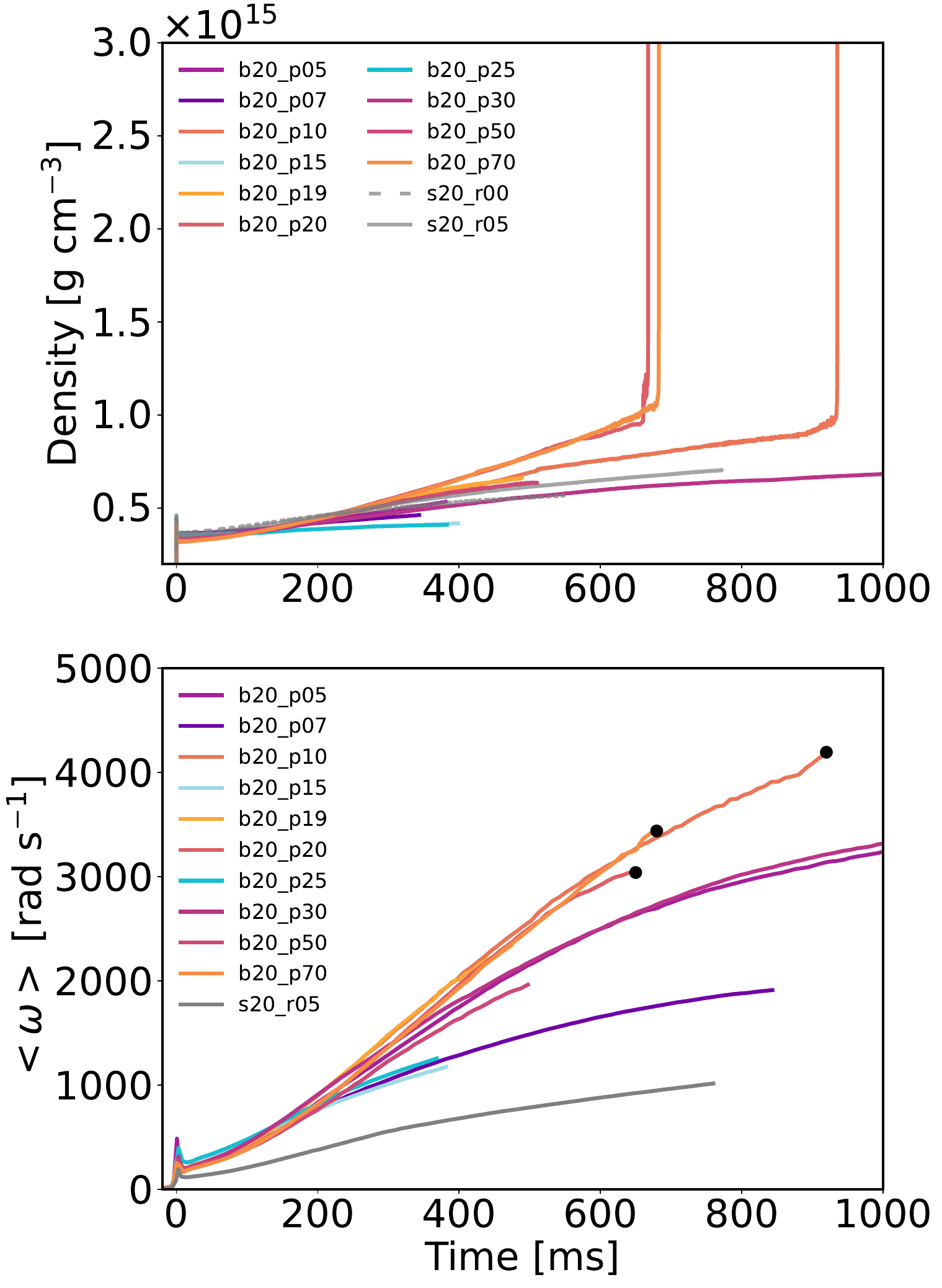}
	\caption{\label{fig:bh}
	\textbf{Upper:} Time evolution of the central density after the core bounce. \textbf{Lower:} Angular velocity evolution of the PNS after the core bounce. Difference colors represent the evolutions from different models in Tabel~\ref{tab:models}. The black solid dots mark the moment of BH formation in our models. }
\end{figure}

Progenitors that fail to explode will ultimately form BHs. 
Figure~\ref{fig:bh} shows the central density evolution of our 12 models. 
We define a BH is formed when the PNS central density rapidly increases, reaching the density limit of the EOS table ($\rho_{\rm max}\sim3 \times 10^{15}$ g cm$^{-3}$).
This can be seen as the second collapse at around 600-1000~ms postbounce in Figure~\ref{fig:bh}. 
For successful explosion models, the shock radius expands so rapidly that it becomes impractical to continue running the simulation due to the high computational costs involved. In addition, some PNS undergo slow mass accretion and, therefore, do not form a BH within our considered simulation time. 
There are three binary progenitors that reach BH formation during our simulation (models {\tt b20\_p10}, {\tt b20\_p20} and {\tt b20\_p70}). 
The corresponding BH formation times for these progenitors are 987~ms, 667~ms, and 699~ms, respectively (see Table~\ref{tab:bh}). 
The lower panel of Figure~\ref{fig:bh} illustrates the angular velocity evolution of PNS in our models. We could see that the angular velocity steadily increases from approximately $\sim 250$~rad~s$^{-1}$ to over $\sim 3000$~rad~s$^{-1}$ after the core bounce, attributed to the accretion of angular momentum and the cooling of the PNS. For the models with high compactness parameters ({\tt b20\_p10}, {\tt b20\_p20} and {\tt b20\_p70}), they spin up with a faster rate due to higher mass accretion rates and eventually form BHs.

\begin{deluxetable}{cccc}
\tablewidth{0pt} 
\tablenum{2}
\label{tab:bh}
\tablecaption{Physical properties at BH formation: 
the BH formation delay time, $t_{BH}$, 
the BH baryonic mass, $M_{BH}$, and the BH spin parameter, $a$}

\tablehead{
Model Name & $t_{\rm BH}$ [ms] & $M_{\rm BH}$ [$M_\odot$] &
$a$ }
\startdata 
b20\_p10  & $987$ & $3.05$  & $0.424$ \\
b20\_p20  & $667$ & $2.89$  & $0.351$ \\
b20\_p70  & $699$ & $2.85$  & $0.335$ \\
\hline
\enddata
\end{deluxetable}

\begin{figure*}
        \plottwo{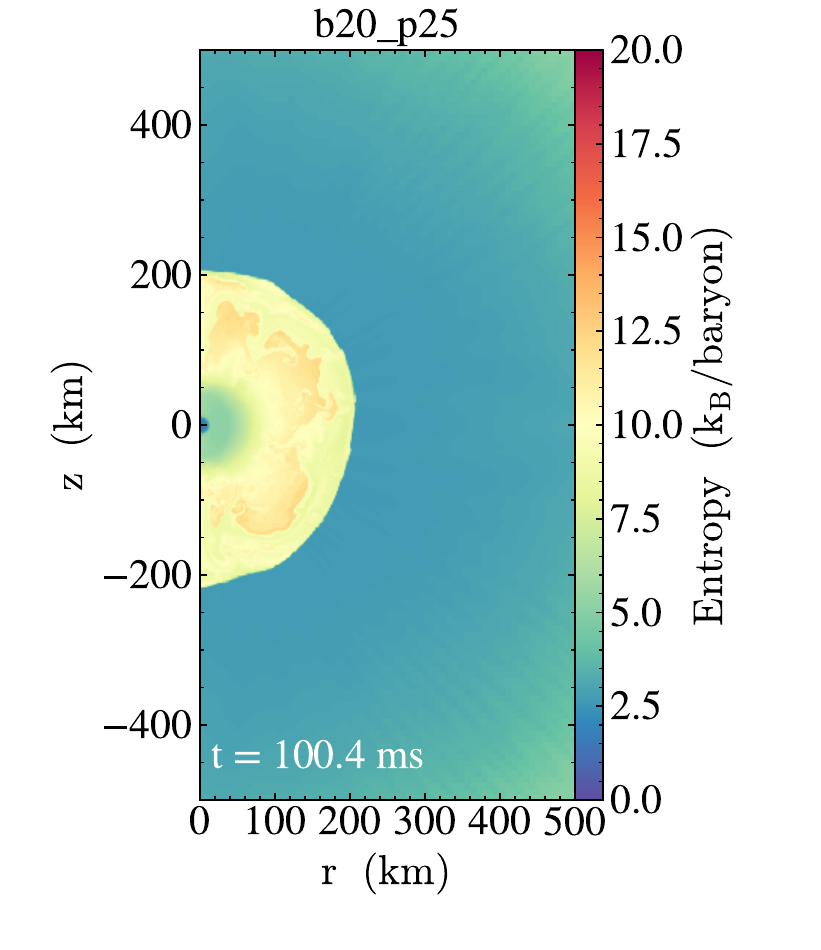}{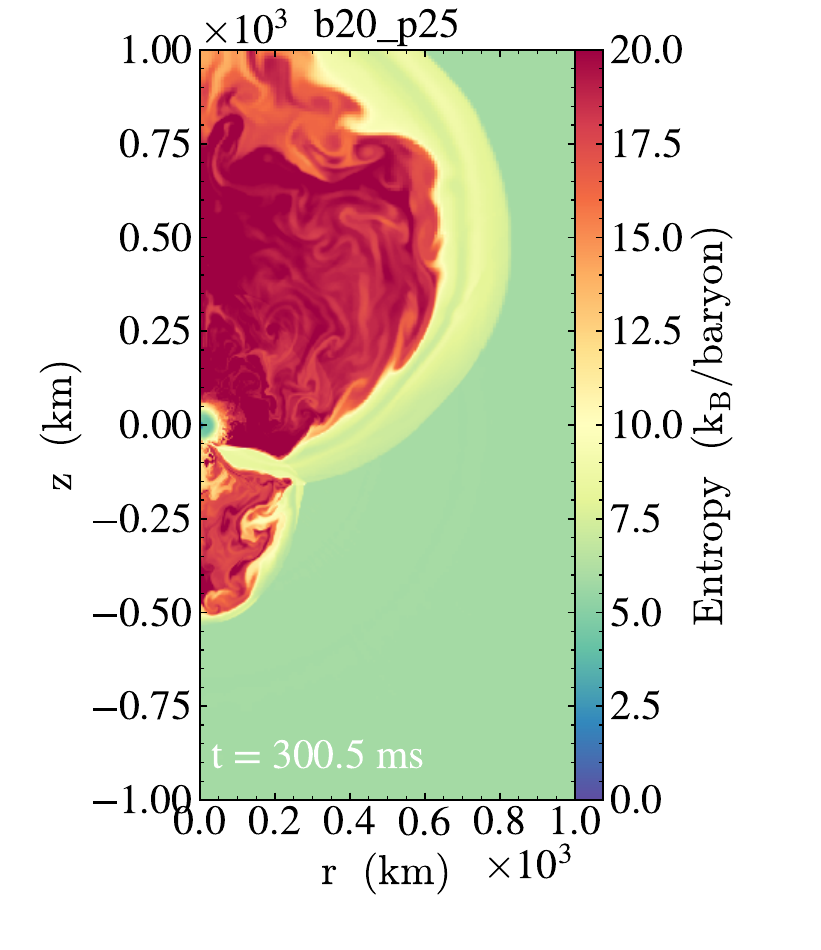}
        \plottwo{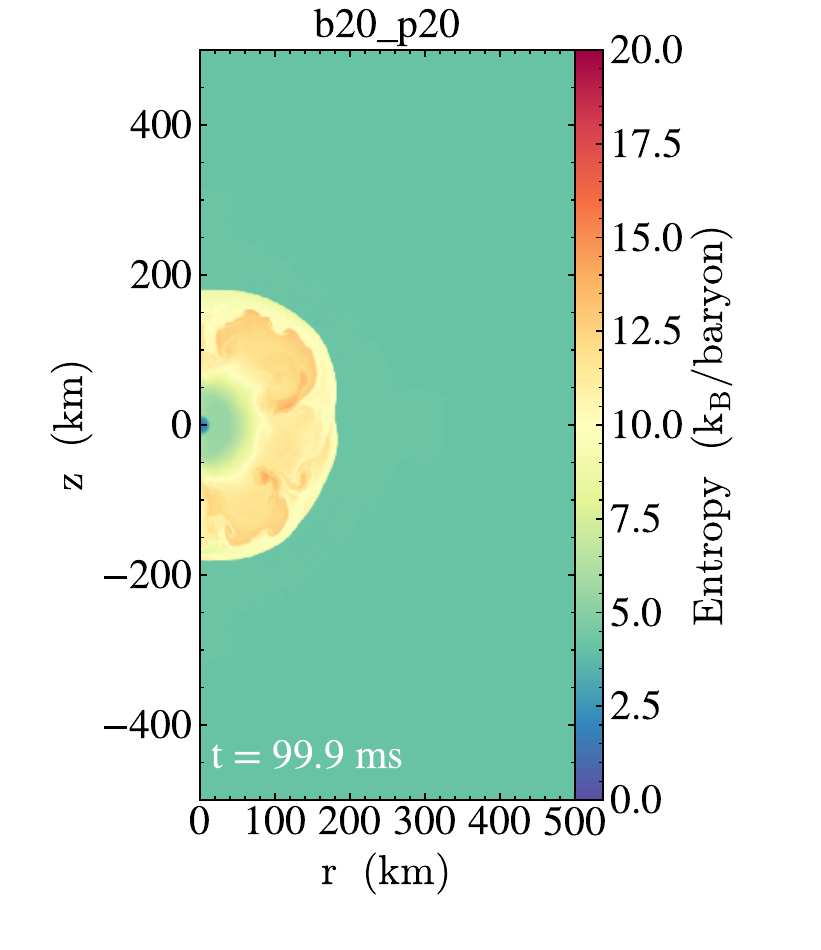}{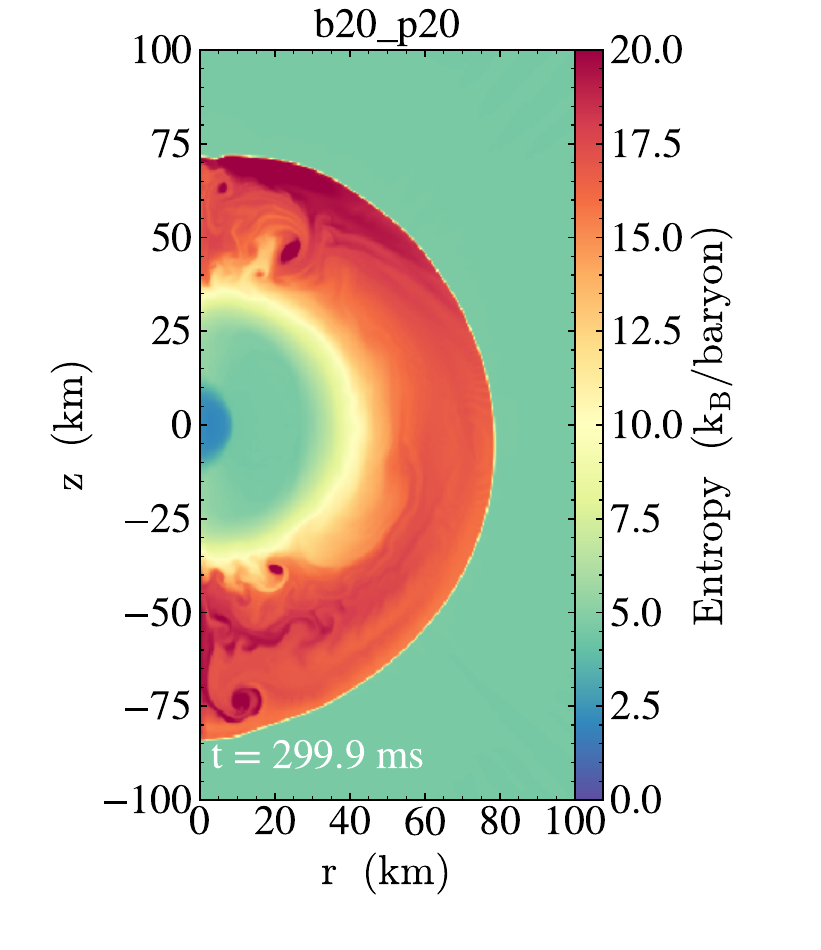}
	\caption{\label{fig:slice}
	The entropy distribution at 100~ms (left) and 300~ms (right) post-bounce. The upper panels show a typical successful CCSN model ({\tt b20\_p25}), while the lower panels are a typical failed CCSN model ({\tt b20\_p20}). }
\end{figure*}

If we assume the specific angular momentum is conserved during the BH formation, then we can estimate the BH spin parameter by computing the PNS mass and angular momentum ($a=J/M$), the BH properties are summarized in Table~\ref{tab:bh}. 
In addition, we observe that the longer the delay time, the larger the resulting PNS baryonic mass before BH formation (see Table~\ref{tab:bh}), even with the same nuclear Equation of State. 
It is important to note that the spin parameters we estimated here may be underestimated in our 2D simulations. As pointed out in \cite{2021ApJ...914..140P} (and the references therein), the spiral mode of SASI could further redistribute the angular momentum and, therefore, potentially lead to further delay of the BH formation and enlarge the spin of the BH. 

Figure~\ref{fig:slice} shows the entropy distribution of two typical CCSN simulations at 100~ms (left) and 300~ms (right) postbounce.
The upper panels present a typical exploding model, {\tt b20\_p25}, while the lower panels show a typical failed supernova model, {\tt b20\_p20}.  
The Exploding progenitor model in the upper panels displays a strong asymmetry explosion, where the ejecta is concentrated in the northern hemisphere, while the mass accretion continues in the southern hemisphere. Other exploding models share a similar evolution.   
This feature is also similar to what has been reported in \cite{2023arXiv230805798B} with a 3D simulation.
On the other hand, models leading to BH formation depict stalled and shrinking shock waves due to strong infall. 
Furthermore, these failed supernovae exhibit stronger SASI activities when compared to successful explosive models. This phenomenon will be reflected in the GW emissions as well (see Section~\ref{sec:mult}).

\subsection{CCSN multimessenger signals}
\label{sec:mult}

\begin{figure}
	\plotone{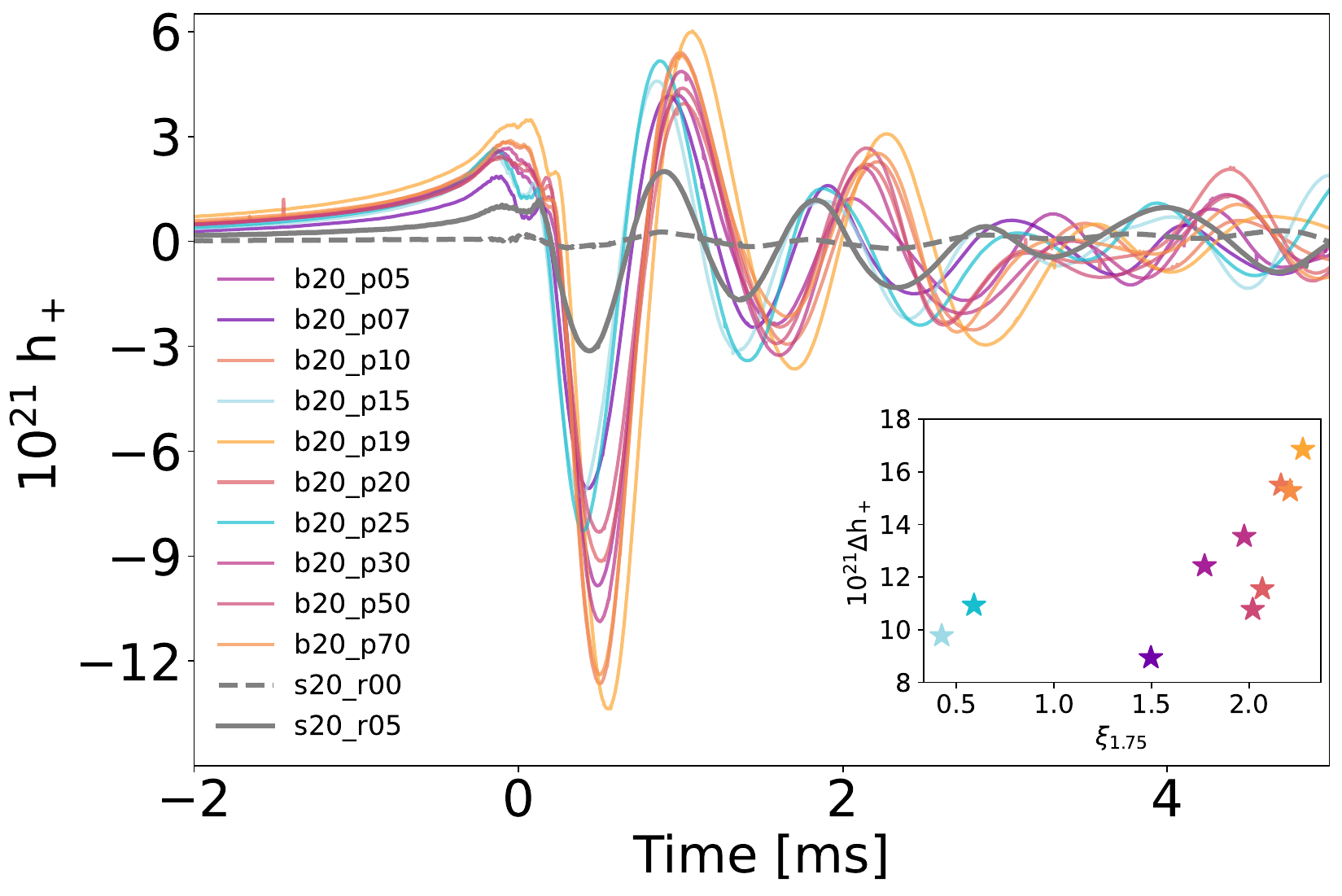}
	\caption{\label{fig:gw_evo}
    The time-domain GW waveforms from the bounce signal. Colored lines represent GW waveforms from binary progenitors, while gray lines show the GW waveforms from single-star progenitors. The embedded subfigure describes the correlations between the compactness parameters and the maximum strain difference within the bounce signals. }
\end{figure}

Figure \ref{fig:gw_evo} presents the time-domain GW waveforms from the bounce signals for each model, accompanied by an embedded subfigure showing the correlation between the core compactness parameter and the maximum GW strain difference within the bounce signals (defined as $\Delta h_+$).
It is evident that our binary progenitors exhibit significantly stronger bounce signals compared to the single-star progenitors, even when the single-star progenitor had an initial rotation close to the critical velocity (model {\tt s20\_r05}). 

Compared to previous simulations with artificial rotational profiles, our binary progenitor models provide equivalent $\Delta h_+$ as the rapid rotating models in \cite{2017PhRvD..95f3019R, 2021ApJ...914...80P}.
Note that \cite{2021ApJ...914...80P} also conducted CCSN simulations with a $20M_\odot$ progenitor model (s20) from \cite{2007PhR...442..269W} using artificial rotations. However, we cannot directly compare the $\Delta h_+$ or the initial rotational speed ($\Omega_0$ defined in their paper) used in \cite{2007PhR...442..269W} to the values in our binary progenitor models. 
This is due to significant differences in the final progenitor mass and the central density at collapse, even though the iron-core masses are similar. 
Nevertheless, we find that the maximum strength of the bounce signal (see Figure~\ref{fig:gw_evo}) and later on rotational speed evolution (see Figure~\ref{fig:bh}) in our models support self-consistent fast-rotating CCSN progenitors can be formed through binary evolution and might emit robust GW bounce signals.

\begin{figure}
	\plotone{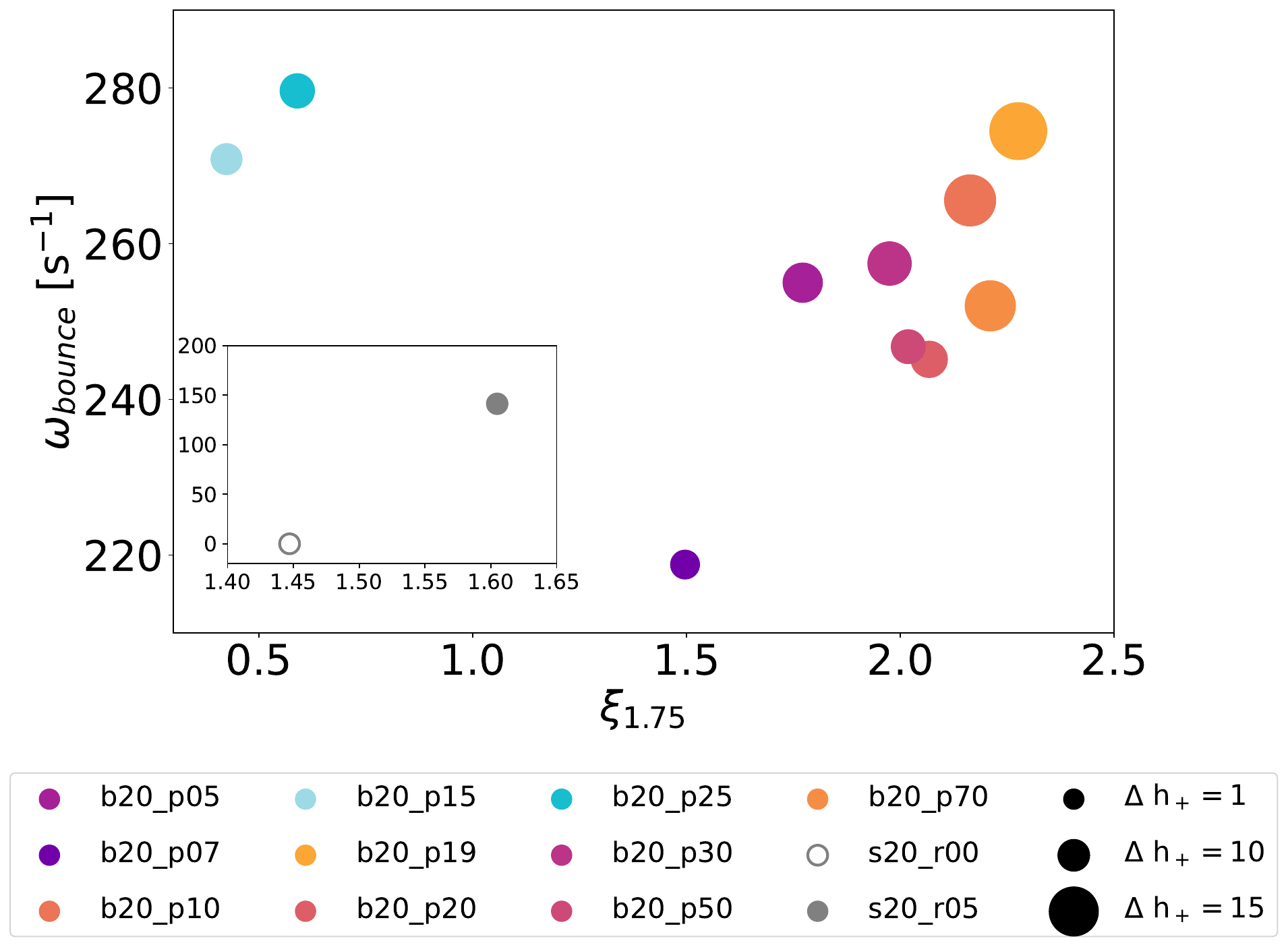}
	\caption{\label{fig:dh}
    The correlations between the PNS bounce angular velocity ($\omega_{\rm bounce}$), the compactness parameter ($\xi_{1.75}$) at $\rho_c = 10^{11}$ g cm$^{-3}$, and the GW strains ($\Delta{h_+}$). The values of $\Delta{h_+}$ are demonstrated by the size of the dots. The embedded subfigure shows the values from single-star models. }
\end{figure}

As mentioned in Section \ref{sec:binary}, there is no obvious correlation between the initial orbital period and the final progenitor structure. GW bounce signals also show no correlation with the initial period. Notably, there is a wealth of work discussing the impact of various progenitor properties on GW signals. 
For instance, \cite{2013ApJ...766...43M} illustrated that larger progenitor sizes form stronger GW signals.  
\cite{2017PhRvD..95f3019R} and \cite{2021ApJ...914...80P} explored the investigation of the positive relation between rotating speed and GW signals.

Consequently, our focus shifts to identifying specific progenitor properties that influence GW signals. In the subplot of Figure \ref{fig:gw_evo}, we compare the compactness parameter and $\Delta{h_+}$ in each binary model. We observe that, within each model, the strength of the bounce signals exhibits a positive correlation with the compactness parameter. 
Furthermore, we investigate the correlation between $\Delta h_+$, the compactness parameter, and bounce PNS angular velocity in Figure~\ref{fig:dh}. The size of the solid dots represents the value of $\Delta h_+$. It can be observed that the models with a larger compactness parameter and faster rotating speed, located in the upper-right region, generate a stronger GW bounce signal ($\Delta h_+$). 
Moreover, single-star models in the subfigure show that the strength of the bounce signal is considerably lower compared to binary progenitors.

\begin{figure*}
	\plotone{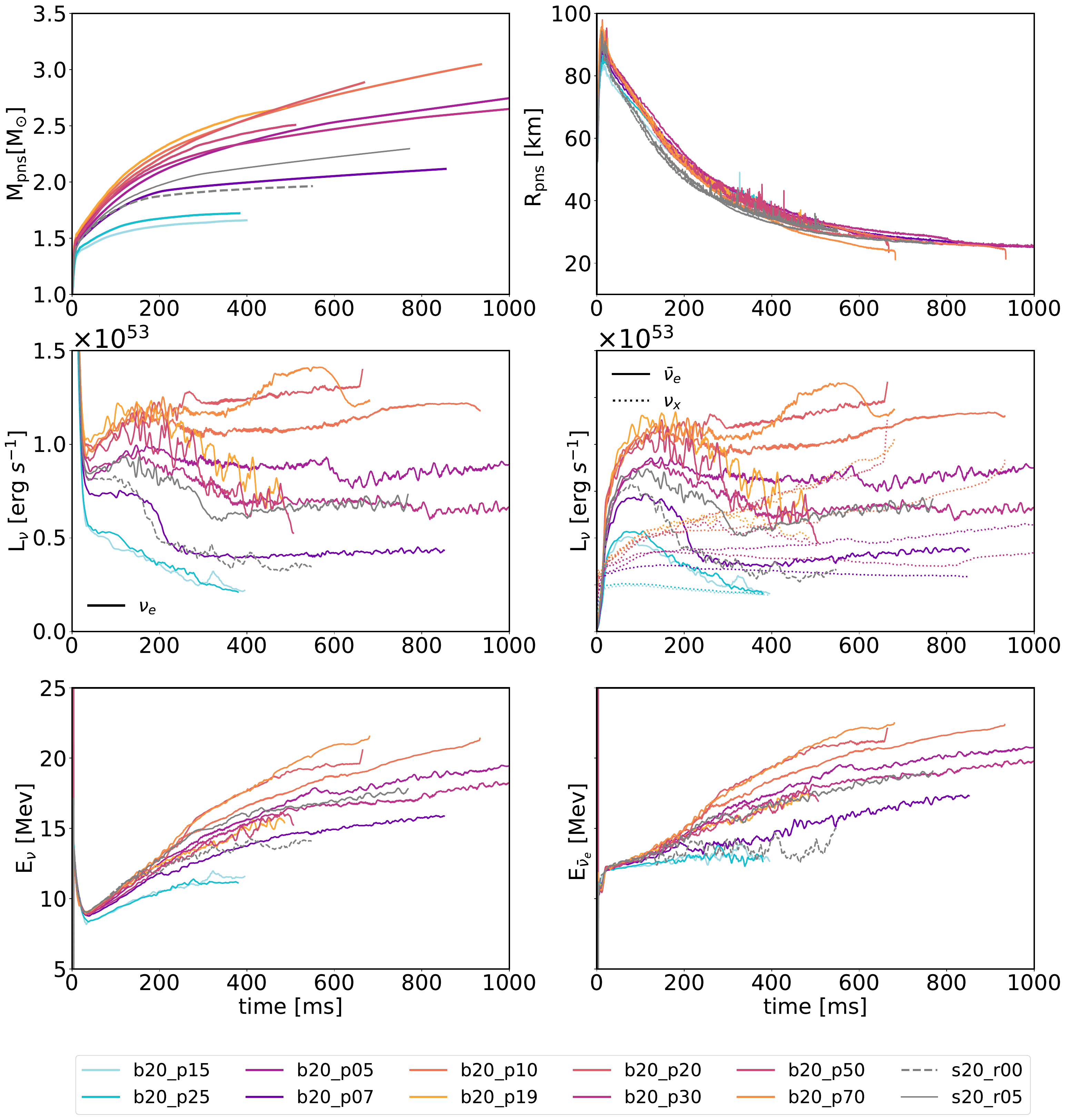}
	\caption{\label{fig:time_evo}
	Time evolution of the PNS mass ($M_{\rm{pns}}$ in the upper left panel), PNS radius ($R_{\rm{pns}}$; upper right), electron-type neutrino luminosity ($L_\nu$; middle left), electron anti-neutrino and heave neutrino luminosities (middle right), electron neutrino mean energy (lower left), and electron anti-neutrino mean energy (lower right). Different colors represent different progenitor models in Tabel~\ref{tab:models}.}
\end{figure*}
\begin{figure*}
	\epsscale{1.2}
        \plotone{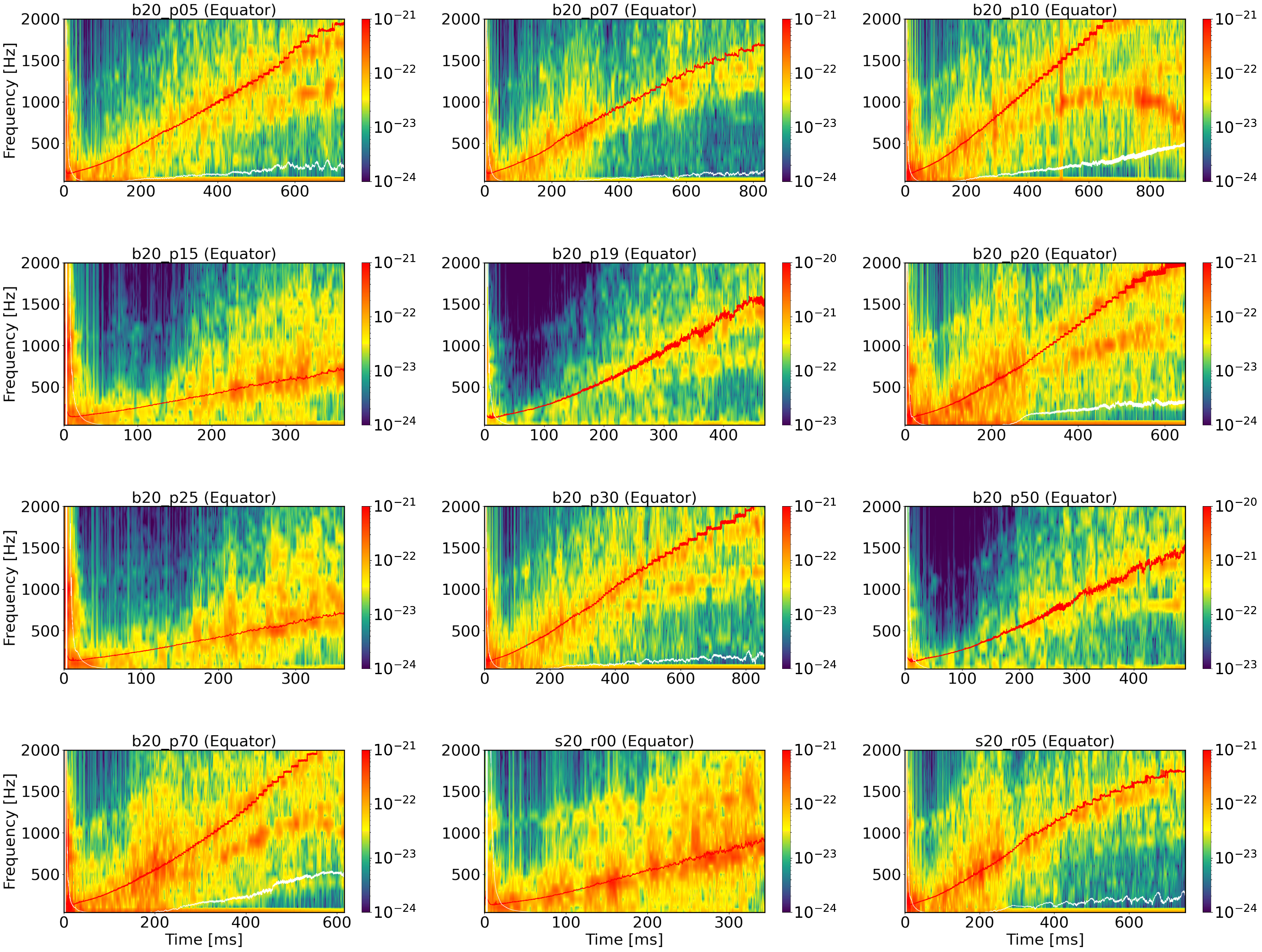}
	\caption{\label{fig:spec}
    GW spectrograms with different progenitor models. The red solid lines represent the GW peak frequencies from the PNS oscillation, and the white solid lines represent the GW SASI frequencies. }
\end{figure*}
\begin{figure*}
    \plotone{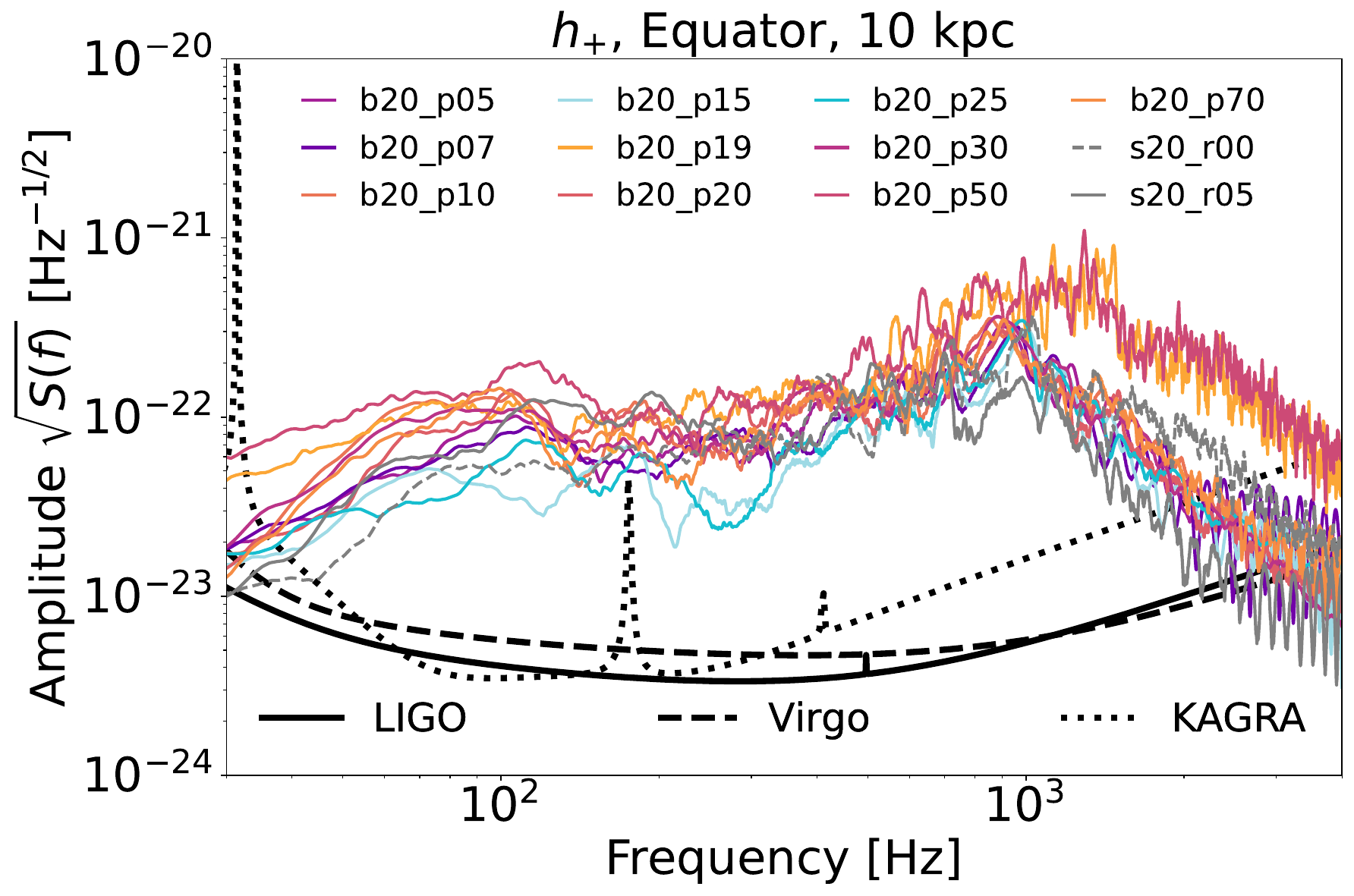}
    \caption{\label{fig:asd}
    GW ASD diagram of our simulated models, assuming a distant $d=10$~kpc. Different colors represent the ASD from different progenitor models in Tabel~\ref{tab:models}. The gray lines show the designed sensitivity curves of the advanced LIGO, Virgo, KAGRA.}
\end{figure*}

As emphasized in Section~\ref{sec:intro}, multimessenger signals are crucial for investigating the inner physics of CCSN.
Figure \ref{fig:time_evo} presents the time evolution of PNS properties, neutrino luminosity, and neutrino mean energy for different progenitor models. 
These PNS properties closely correlate with the mass accretion rate, with higher compactness parameters resulting in more substantial mass accretion rates \citep{2009A&A...499....1F,2015PASJ...67..107N}. 
Notably, models forming BHs earlier exhibit larger mass accretion rates, leading to the formation of heavier PNSs. 
Conversely, models with lower compactness cores, such as {\tt b20\_p15} and {\tt b20\_p25}, feature lower mass accretion rates, resulting in both lighter PNSs and more readily achievable CCSN explosions. 

The middle and lower panels of Figure \ref{fig:time_evo} display the time evolution of neutrino luminosity and neutrino mean energy. 
As discussed in \cite{2013ApJ...762..126O}, the (anti-)neutrino luminosity and energy are strongly correlated with the compactness parameter. These emissions are influenced by the mass accretion rate. 
Our simulations illustrate that higher compactness progenitors yield greater mass accretion rates, consequently leading to more substantial (anti-)neutrino luminosity and higher neutrino energy emissions. Similar neutrino emission behavior can also be found in other literature \citep{2009A&A...499....1F,2015PASJ...67..107N,2018ApJ...855L...3O}. 
 
In order to search for the GW events from CCSNe, the gravitational waveform templates from CCSN simulations provide useful information for the GW observations. 
In Figure \ref{fig:spec}, we present the GW spectrogram from our models. \cite{2013ApJ...766...43M,2019PhRvL.123e1102T,2021PhRvD.104l3009S} have provided semi-analytical descriptions to estimate the PNS peak frequency. In this work, we follow the method in \cite{2018ApJ...857...13P} to modify the equation in \cite{2013ApJ...766...43M} and find that the peak PNS frequency can be better matched by 


\begin{equation}
f_{\rm peak} = \frac{1}{2\pi} 
    \frac{G M_{\rm PNS} }{R^2_{\rm PNS} c} 
    \sqrt{ 1.1
        \frac{M_n}{\left< E_{\bar{\nu}_e} \right> }
    },\label{eq:gmode}
\end{equation}

where $M_{\rm PNS}$ is the mass of the PNS, 
$R_{\rm PNS}$ is the radius of the PNS, 
$M_n$ is the proton mass, 
$c$ is the light speed, 
and $\left< E_{\bar{\nu}_e} \right>$ is the mean energy of electron anti-neutrinos \citep{2013ApJ...766...43M, 2018ApJ...857...13P}.

In addition, we follow the equations derived from \cite{2014ApJ...788...82M, 2021MNRAS.503.2108P} to approximate the SASI frequency,
\begin{equation}
    f_{\rm SASI} = \frac{1}{12~{\rm ms}} \left(\frac{R_{\rm sh}}{100~{\rm km}} \right)^{-3/2} \left[\ln \left( \frac{R_{\rm sh}}{R_{\rm PNS}}\right) \right]^{-1}, \label{eq:sasi}
\end{equation}
where the $R_{\rm sh}$ is the averaged shock radius.

Note that the fitting parameters we used here are slightly different from the equations provided by \cite{2013ApJ...766...43M} and \cite{2014ApJ...788...82M}. 
We find that Equations~\ref{eq:gmode} and \ref{eq:sasi} can better describe the GW features in our CCSN simulations. These differences in the fitting parameters may originate from the different gravity or neutrino treatments in our CCSN code. 

All models show very strong PNS surface oscillations (the red lines), 
but the SASI components (the white lines) are only visible in models {\tt b20\_p10}, {\tt b20\_p20} and {\tt b20\_p70}, which evolved much longer than other models and eventually experienced BH formation. 
In addition, a higher-order g-mode signal can be seen between the red and white line after $\sim 400$~ms postbounce. Interestingly, this higher-order mode signal in certain models, such as {\tt b20\_p05}, {\tt b20\_p10}, {\tt b20\_p20}, {\tt b20\_p30}, and {\tt b20\_p70}, seem to decline at very late times, prior to BH formation. 

Figure~\ref{fig:asd} shows the amplitude spectral density (ASD) of our simulated waveforms for an observer at a distance $d=10$~kpc from the CCSN, viewing from the orbital plane. 
Unsurprisingly, all our models are possible to be detected by the current LVK network, but we point out that GW signals from CCSNe tend to prefer kilo-hertz window \citep{2018ApJ...857...13P}.
In addition, in the hundred-hertz window, the ASD amplitude is proportional to the compactness parameter. In the kilo-hertz window, the two exploding models ({\tt b20\_p19} and {\tt b20\_p50}) show extremely loud high-frequency GW signals due to asymmetric explosions.  

\section{Summary \& Conclusions}
\label{sec:summary}

Fast-rotating CCSN progenitors are expected to produce stronger gravitational wave signals compared to their non-rotating counterparts \citep{2017PhRvD..95f3019R,2018MNRAS.475L..91T,2019MNRAS.486.2238A,2019ApJ...878...13P,2020MNRAS.494.4665P,2020MNRAS.493L.138S,2021ApJ...914..140P, 2022MNRAS.510.5535J, 2023arXiv231020411H}. However, the formation of rapidly rotating CCSN progenitors poses a challenging task within the framework of single-star stellar evolution. In this study, we conducted binary stellar evolution from ZAMS to CCSN progenitor, generating a sequence of realistically rapid-rotating progenitors. Subsequently, we performed self-consistent CCSN simulations to investigate the dynamical evolution of CCSNe and their multimessenger signals.

Our exploration of stellar evolution within binary systems revealed that there is no linear correlation between the final progenitor structure and the initial orbital period. However, during the course of stellar evolution, specific accretors ({\tt b20\_p15} and {\tt b20\_p25}) underwent strong convection during helium burning, resulting in significantly lower carbon fractions but higher oxygen fractions. These structural differences affect the subsequent evolution and progenitor structure. Models with strong convection exhibited the merger of silicon and oxygen shells, forming a smaller iron core, resulting in a significantly lower compactness parameter. Additionally, shell merging reduced angular momentum transport efficiency, resulting in a faster-rotating inner core. In our research, binary progenitors gave rise to fast-rotating CCSN progenitors, comparable to the extremely fast-rotating progenitors in \cite{2021ApJ...914...80P}.

On the other hand, other accretors with weaker convection formed distinct boundaries between different shells. These accretors formed heavier iron cores with smaller density gradients at the iron-silicon shell interface, then made CCSN progenitors with higher compactness parameters compared to single-star progenitors and strong convection models.

Furthermore, we observed an inverse correlation between the bounce density and the compactness parameter (see Figure~\ref{fig:bc_dens}), aligning with the relationship discussed in \cite{2002RvMP...74.1015W}. We also observed that there are several physical properties that determine the outcome of CCSN, including the compactness parameter, progenitor density gradient, angular velocity, and mass accretion rate. In our simulations, three early BH formation models ({\tt b20\_p10}, {\tt b20\_p20} and {\tt b20\_p70}) exhibited robust mass accretion, resulting in the formation of rapidly rotating stellar mass BH.

Regarding multimessenger signals in our simulations, 
we have identified a positive correlation between the strength of GW bounce signals and the compactness parameter and angular velocity. 
Moreover, the properties of PNS and neutrino emission are strongly influenced by the mass accretion rate, which is also linked to the compactness parameter.

We have demonstrated all our fast-rotating models show strong PNS surface oscillations, and some of them also display higher-order mode signals.
In addition, the models leading to BH formation (see Tabel~\ref{tab:bh}) exhibit weak SASI signals prior to BH formation, and their higher-order mode signals show a decline in GW frequency at late time. 

Lastly, our simulations have indicated that GW emissions are detectable by the LVK observatories if the sources are located at around 10~kpc. Additionally, progenitors with higher compactness parameters can emit stronger GW signals in the hundred-hertz window, and the exploding fast-rotating binary progenitor models show significantly higher GW signatures in the kilo-hertz window. This presents a slightly more optimistic perspective for the future GW detection from CCSNe when considering binary progenitors.

While our 2D simulations allow for numerous CCSN simulations, they have limitations in terms of incorporating more robust microphysics. In an ideal scenario, conducting 3D simulations, including full GR treatments, would be preferable. For instance, \cite{2021MNRAS.508..966T}, \cite{2023MNRAS.520.5622B}, and \cite{2023arXiv231020411H} recently emphasized that low-$T/|W|$ instability leads to a strong burst in GW emission, where the $l=1$ modes are prohibited in 2D simulations.
Moreover, \cite{2017MNRAS.468.2032A} pointed out that the amplitudes of high-frequency components resulting from PNS surface oscillation are significantly lower in 3D than in 2D. Additionally, our simulations employed an effective General Relativity (GR) potential to approximate GR effects, limiting our ability to fully simulate BH formation. Recent studies by \cite{2023MNRAS.526..152K} have conducted full relativistic simulations. Thus, we remind the readers that the effects from three-dimensional fluid instabilities, full GR, and magneto-hydrodynamics are not included in this study and will be our future work. 

\acknowledgments
  
This work is supported by the National Science and Technology Council of Taiwan through grants MOST 110-2813-C-007-013-M, NSTC 111-2112-M-007-037 and 112-2112-M-007-040,
by the Center for Informatics and Computation in Astronomy (CICA) at National Tsing Hua University through a grant from the Ministry of Education of Taiwan.
{\tt FLASH} was in part developed by the DOE NNSA-ASC OASCR Flash Center at the University of Chicago.
The simulations and data analysis have been carried out at the {\tt Taiwania~3} supercomputer in the National Center for High-Performance Computing (NCHC) in Taiwan, 
and on the CICA cluster at National Tsing Hua University.
Analysis and visualization of simulation data were completed using the analysis toolkit {\tt yt}.
\software{MESA \citep{2011ApJS..192....3P}, FLASH \citep{2000ApJS..131..273F, 2008PhST..132a4046D}, yt \citep{2011ApJS..192....9T}, Matplotlib \citep{2007CSE.....9...90H}, NumPy \citep{2011CSE....13b..22V}, SciPy \citep{2019zndo...3533894V}}


\end{document}